\documentclass[fleqn,usenatbib]{mnras}

\usepackage[T1]{fontenc}

\DeclareRobustCommand{\VAN}[3]{#2}
\let\VANthebibliography\thebibliography
\def\thebibliography{\DeclareRobustCommand{\VAN}[3]{##3}\VANthebibliography}

\usepackage{graphicx}	
\usepackage{amsmath}	
\usepackage{amssymb}	
\usepackage{xcolor}
\usepackage{cancel}
\usepackage{newtxtext,newtxmath}

\title[SN 2020znr Polarimetry]{Post maximum light and late time optical imaging polarimetry of type I superluminous supernova 2020znr}

\author[F. Poidevin et al.]{F. Poidevin$^{1,2}$,\thanks{E-mail: fpoidevin@iac.es (IAC)}
C. M. B. Omand$^{3}$,
  I. P\'{e}rez-Fournon$^{1,2}$,
  R. Clavero$^{1,2}$,
  R. Shirley$^{4,5}$,
  R. Marques-Chaves$^{6}$,
  \newauthor 
C. Jimenez Angel$^{1,2}$,
S. Geier$^{1,7}$.
\\
$^{1}$ Instituto de Astrofis\'{i}ca de Canarias, 38200 La Laguna,Tenerife, Canary Islands, Spain\\
$^{2}$ Departamento de Astrof\'{\i}sica, Universidad de La Laguna (ULL), 38206 La Laguna, Tenerife, Spain\\
$^{3}$ The Oskar Klein Centre, Department of Astronomy, Stockholm University, AlbaNova, SE-10691 Stockholm, Sweden\\
$^{4}$Astronomy Centre, Department of Physics \& Astronomy, University of Southampton, Southampton, SO17 1BJ, UK\\
$^{5}$Institute of Astronomy, University of Cambridge, Madingley Road, Cambridge, CB3 0HA, UK\\
$^{6}$Geneva Observatory, University of Geneva, Chemin Pegasi 51, CH-1290 Versoix, Switzerland\\
$^{7}$ GRANTECAN: Cuesta de San Jos\'{e} s/n, 38712 Bre{\~n}a Baja, La Palma, Spain\\
}

\date{Accepted XXX. Received YYY; in original form ZZZ}

\pubyear{2021}

\begin{document}
\label{firstpage}
\pagerange{\pageref{firstpage}--\pageref{lastpage}}
\maketitle

\begin{abstract}
Optical imaging polarimetry was conducted on the hydrogen poor superluminous supernova 2020znr during 3 phases after maximum light ($\approx$ +34 days, +288 days and +289 days). After instrumental and interstellar polarization correction, all measurements are consistent with null-polarization detection. Modelling the light curve with a magnetar spin-down model shows that SN2020znr has similar magnetar and ejecta parameters to other SLSNe.
A comparison of the best-fit values discussed in the literature on SN 2017egm and SN 2015bn, two hydrogen poor SLSNe showing an increase of polarization after maximum light, suggests that SN 2020znr has higher mass ejecta that may prevent access to the geometry of the inner ejecta with optical polarimetry. The combined information provided by spectroscopy and light curve analysis of type I SLSNe may be an interesting avenue to categorize the polarization properties of this class of transients. This approach would require to expand the sample of SLSNe polarimetry data currently available with early and late time epochs new measurements.  
\end{abstract}

\begin{keywords}
techniques: polarimetric – supernovae: general – supernovae: individual: 2020znr, 2015bn, 2017egm, 2018hti, LSQ14mo, PTF12dam.
\end{keywords}



\section{Introduction} \label{intro}

Superluminous supernovae (SLSNe) are core collapse supernovae
showing explosions about 100 times brighter than
common supernovae. While their population is quite rare \citep[e.g.][]{schulze2021} based on current survey detection rates, and limited resources to classify all of them with spectroscopy \citep[][]{nicholl2021},
they have been subject of intense studies since their discovery
about 15 years ago \citep[][]{2019ARA&A..57..305G}. This transient population is mainly subdivided in two classes, which are hydrogen-poor (SLSN-I) and hydrogen-rich (SLSN-II). Their light-curves show a disparity of shapes, and of rising and decreasing time scales,  \citep[][]{2017ApJ...850...55N, 2018ApJ...852...81L, 2020ApJ...901...61L}, likely subject to local and global environmental effects \citep[][]{2016ApJ...830...13P, 2018MNRAS.473.1258S, schulze2021}, making their classification \citep[][]{2020ApJ...904...74G, 2021AJ....161..141S} and interpretation quite complex.

Several models were proposed to explain SLSNe.  The three main models are the pair-instability model, the ejecta-circumstellar material interaction model, and the magnetar spin-down model \citep{2018SSRv..214...59M, 2019ARA&A..57..305G, 2021arXiv210908697N}. Pair instability supernovae are the explosions of extremely massive ($\gtrsim$ 130 $M_\odot$) stars \citep{1967ApJ...148..803R, 1967PhRvL..18..379B, 1968Ap&SS...2...96F}; the light curves of these explosion can be consistent with some slower SLSNe \citep{2016ApJ...831..144L, 2018MNRAS.479.3106K, 2018ApJ...860..100D}, but their spectra are predicted to be red and spectral models are inconsistent with observations in both the early and nebular phases \citep{2012MNRAS.426L..76D, 2016MNRAS.455.3207J, 2017ApJ...835...13J}. Ejecta-CSM interaction, where the supernova ejecta collide with material previously ejected from the star, either through a binary interaction, stellar wind, or eruptive mass loss \citep{2014ARA&A..52..487S} such as a pulsational pair instability \citep{2017ApJ...836..244W}, is usually used to explain SLSNe-II \citep{2010ApJ...709..856S, 2019ARA&A..57..305G}, but is also consistent with light curves for SLSNe-I \citep{2017ApJ...835..266T, 2012ApJ...757..178G, 2014MNRAS.444.2096N}.  The magnetar spin-down model, where the rotational energy of a newborn millisecond magnetar is used to power the supernova \citep{og71, kb10, woo10}, can also fit most SLSN-I light curves \citep{2017ApJ...850...55N}.  Ejecta-CSM interaction and magnetar spin-down can be difficult to distinguish through photometry alone, but spectroscopic and multiwavelength follow-up have provided strong candidates for both magnetar powered supernovae \citep{2018ApJ...864L..36M, 2019ApJ...876L..10E} and interaction powered supernovae \citep{2018NatAs...2..887L}, although a hybrid model may be necessary to explain some SLSNe-I \citep{2017MNRAS.468.4642I, 2017ApJ...848....6Y}.

As for many other transients, SLSNe are intensively observed with spectroscopy. Analysis and modelling of time evolving spectral absorption features give access to the photosphere composition and dynamics \citep[e.g.][]{gal-yam2019}. Analysis of large sample spectra database also provides information on possibly different channels leading to such explosions \citep[e.g.][]{2018ApJ...855....2Q, 2021ApJ...909...24K}. Photometry also provides a wealth of information. The higher the sampling rate, the broader the wavelengths bandwidth coverage with multi-band photometry, the better the characterization of the time evolution of the luminosity via intense multicolor light curve modelling \citep[e.g.][]{2017ApJ...850...55N, 2018ApJS..236....6G, kumar2021}.
It can be hard to distinguish various SLSN models using only optical emission, but both magnetar-driven supernovae and interaction-powered supernovae are expected to be bright in radio and X-rays at late times \citep{2015ApJ...805...82M, Kashiyama+16, 2018MNRAS.474..573O}, and various multiwavelength follow-up observations have been also carried out.  \cite{2018ApJ...864...45M} surveyed 26 SLSNe in X-rays at various post-explosion timescales, and only found emission from PTF12dam around its optical peak; X-rays were also detected previously in SCP06F6 \citep{2013ApJ...771..136L}, which was suggested to be due to an engine-powered ionization breakout \citep{2014MNRAS.437..703M}. Radio and millimetre observations have mostly found non-detections \citep{2019ApJ...886...24L, 2021MNRAS.tmp.2267M, 2021ApJ...912...21E, 2021arXiv210810445H}, but three sources have been detected: PTF10hgi \citep{2019ApJ...876L..10E, 2019ApJ...886...24L, 2020MNRAS.498.3863M, 2021ApJ...911L...1H}, which is consistent with the magnetar model; SN 2017ens \citep{2021ATel14393....1C}, which may be either due to a magnetar or CSM shock; and 2020tcw \citep{2021ATel14418....1C}, which was detected only months after explosion and is likely due to CSM interaction.  An infrared excess in magnetar-driven supernova was predicted by \cite{2019MNRAS.484.5468O} due to heating of dust formed in the supernova, and which was observed recently in SN 2018bsz \citep{2021arXiv210907942C}.

Optical polarimetry or spectropolarimetry has been investigated on a relatively small sample of SLSNe \citep[][]{leloudas2015, brown2016, inserra2016, leloudas2017, cikota2018, maund2019, maund2020, maund2021, lee2019, lee2020, cikota2018, saito2020}. Most of the observations were obtained close to peak maximum light, and no significant polarization detection was found ($<3 \sigma$). Up to date, the most stringent detections have been obtained on only two different transients. \citet{inserra2016} and \citet{leloudas2017} discuss null-polarization detection on 2015bn during 4 phases before maximum light but an increase up to about 1.5 $\%$ during later phases (from +5.4 up to +45.8 days). Similarly,  \citet{maund2019} show null-polarization detection during early phases on 2017egm given that the possible polarization signal, of order $1\%$ detected at less than 2$\sigma$ with imaging techniques, was attributed to the spiral arm in the proximity to the position of 2017egm. On the other hand, using spectropolarimetry observations, \citet{saito2020} were able to show that the degree of polarization associated to 2017egm significantly changes from that measured at the earlier phase with an intrinsic polarization of $\sim$ 0.2 $\%$ during early phases and an intrinsic polarization $\sim$ 0.8 $\%$ in the late phase ($+185$ days).

In this work we present the first result, from a ten hour linear 
polarimetry survey, designed to explore the frequency of objects such as SN 2015bn and SN 2017egm. These results were obtained on the
H-poor SN 2020znr during mainly 2 distinct epochs. 
This transient is located at $($RA, Dec$) =
(109.776773^{\circ}, 23.885371^{\circ})$, J2000.
It was discovered by \citet{nordin2020} on November 12, 2020
from ZTF \footnote{Zwicky Transient Facility, {\tt
    https://www.ztf.caltech.edu}.} public alerts \citep[][]{bellm2019}. The discovery
magnitude obtained with the ZTF-cam mounted on the Palomar 1.2
meter Oschin was of 19.77 mag in the g-filter (AB system). 
The transient was later classified as a SLSN-I at a redshift z$=0.1$
by \citet{ihanec2020} from the analysis of a spectrum obtained by
the extended Public ESO Spectroscopic Survey of Transient Objects
\citep[ePESSTO;][]{smartt2017} collaboration.
The spectrum is publicly available on the Transient Name Server
  \footnote{Transient Name Server, {\tt
    https://www.wis-tns.org/2020znr}} (TNS). The transient is very likely associated
to the low brightness feature observed in the Dark Energy Camera Legacy Survey (DECaLS) \footnote{Legacy Survey, {\tt
    https://www.legacysurvey.org/}}, Data
Release 9 (DR9) at position $($RA, Dec$) = (109.7767^{\circ},
23.8854^{\circ})$, J2000, of apparent magnitudes g=23.96, r=23.64, z=23.50 mag, i.e. of absolute magnitudes -14.32, -14.64 and -14.78 mag, respectively, assuming it is at the same redshift as the SLSN at z=0.1.

\section{Observations}

\subsection{Spectroscopy} \label{data_spec}

\begin{figure}
\begin{center}
\vspace*{2mm}
\centering
\hspace*{0.cm}
\includegraphics[width=85mm,angle=0]{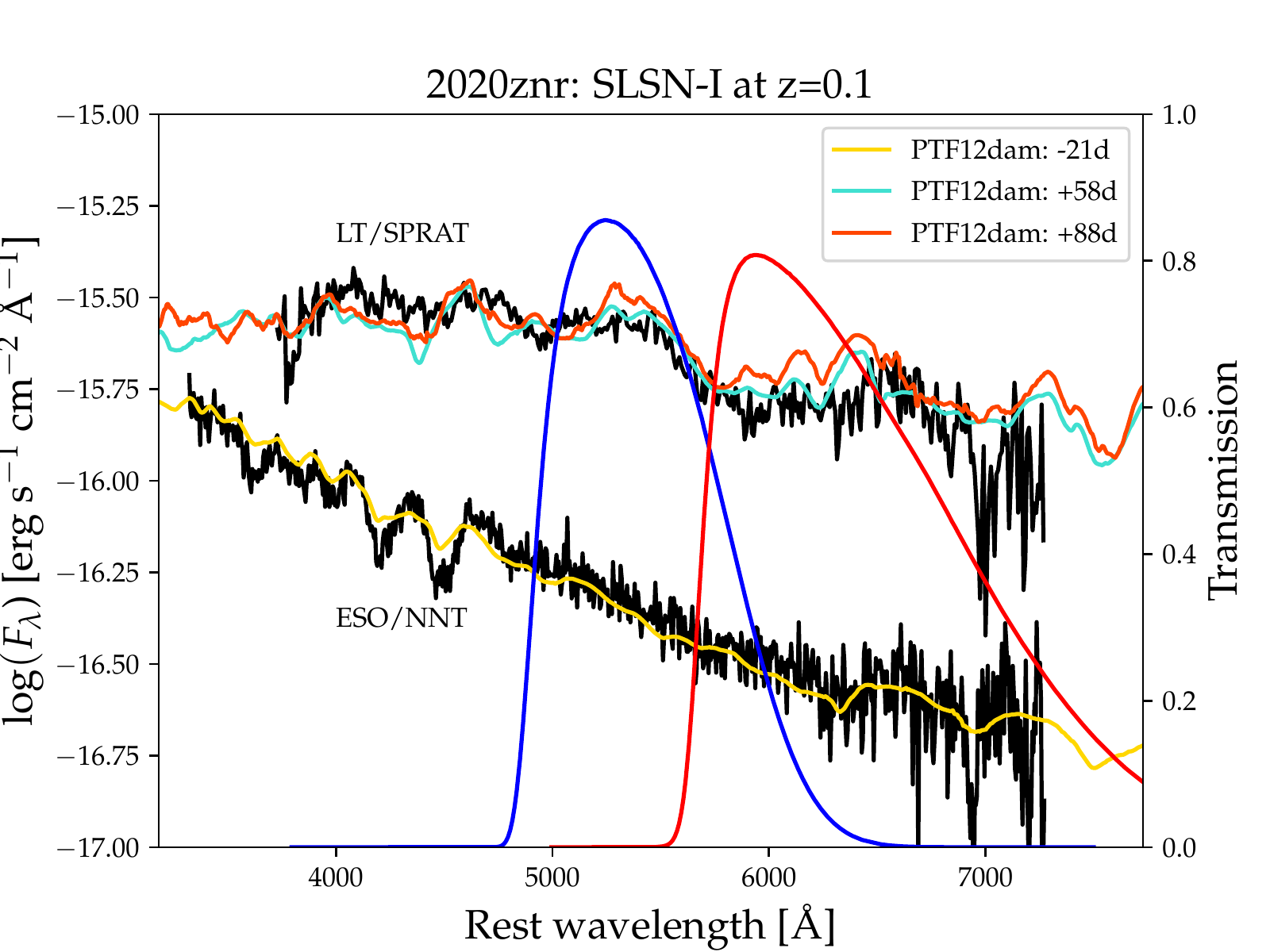}
\vspace*{0.7cm}
\caption{Spectra obtained on SN 2020znr. Plots of spectra of
  PTF12dam obtained at various phases are also displayed for
  comparison. The transmission curves of the V- and R-band filters
  mounted on the ALFOSC are shown with the blue and red curves, respectively.}
\label{fig:2020znr_SPEC_AND_FILT}
\end{center}
\end{figure}

The ePESSTO+ collaboration spectrum of SN 2020znr was obtained with
the ESO-NTT $/$ EFOSC2 on November 17, 2020.
We also obtained a spectrum about 122 days later, on March 19, 2021. This second epoch spectrum was obtained with the Liverpool Telescope (LT) SPectrograph for the Rapid Acquisition of Transients (SPRAT) \citep{2004SPIE.5489..679S,piacscik2014} with a total integration time of
  $3 \times 200 $ seconds through $1.8^{\arcsec}$ with the blue-optimized grating configuration starting at UTC time $=21:38:45.147$.
The two spectra are shown in Figure~\ref{fig:2020znr_SPEC_AND_FILT}.
Also shown in the Figure are the transmission of the V- and
R-band filters mounted on the Alhambra Faint Object Spectrograph and
Camera (ALFOSC) used to get linear polarimetry data discussed in the next section.

The two spectra are compared to other spectra obtained on SLSN-I.
The first epoch spectrum is similar to early epochs spectra showing a blue continuum with absorption features that have been identified as
$O_{\rm II}$ by \citet{quimby2011}. Such features have been further
discussed and modeled by \citet{gal-yam2019}. The simple method
proposed by \citet{gal-yam2019} shows that early phase SLSNe spectra are mainly probing the photosphere which can be mostly described by absorptions from single transitions with a single photospheric velocity. An early phase, 21 days before maximum light, spectrum of PTF12dam is shown in Figure~\ref{fig:2020znr_SPEC_AND_FILT} for comparison. This spectrum was the best match solution from a fitting template analysis with SNID, the SuperNova IDentification code,  
\footnote{{\tt
    https://people.lam.fr/blondin.stephane/software/snid/}} \citep[][]{blondin2007} 
using the \cite{2018ApJ...855....2Q} spectra database ingested in our custom SNID template library.
Later epochs spectra from PTF12dam obtained after maximum light that resemble SN Ic \citep{pastorello2010} spectra  are also found to be relativaly good candidates for comparisons with the LT SPRAT spectrum. From the SNID analysis the best phase estimate of the LT SPRAT spectrum is +100 days. 

\subsection{Photometry} \label{data_phot}

\begin{figure}
\begin{center}
\vspace*{2mm}
\centering
\hspace*{0.cm}
\includegraphics[width=85mm,angle=0]{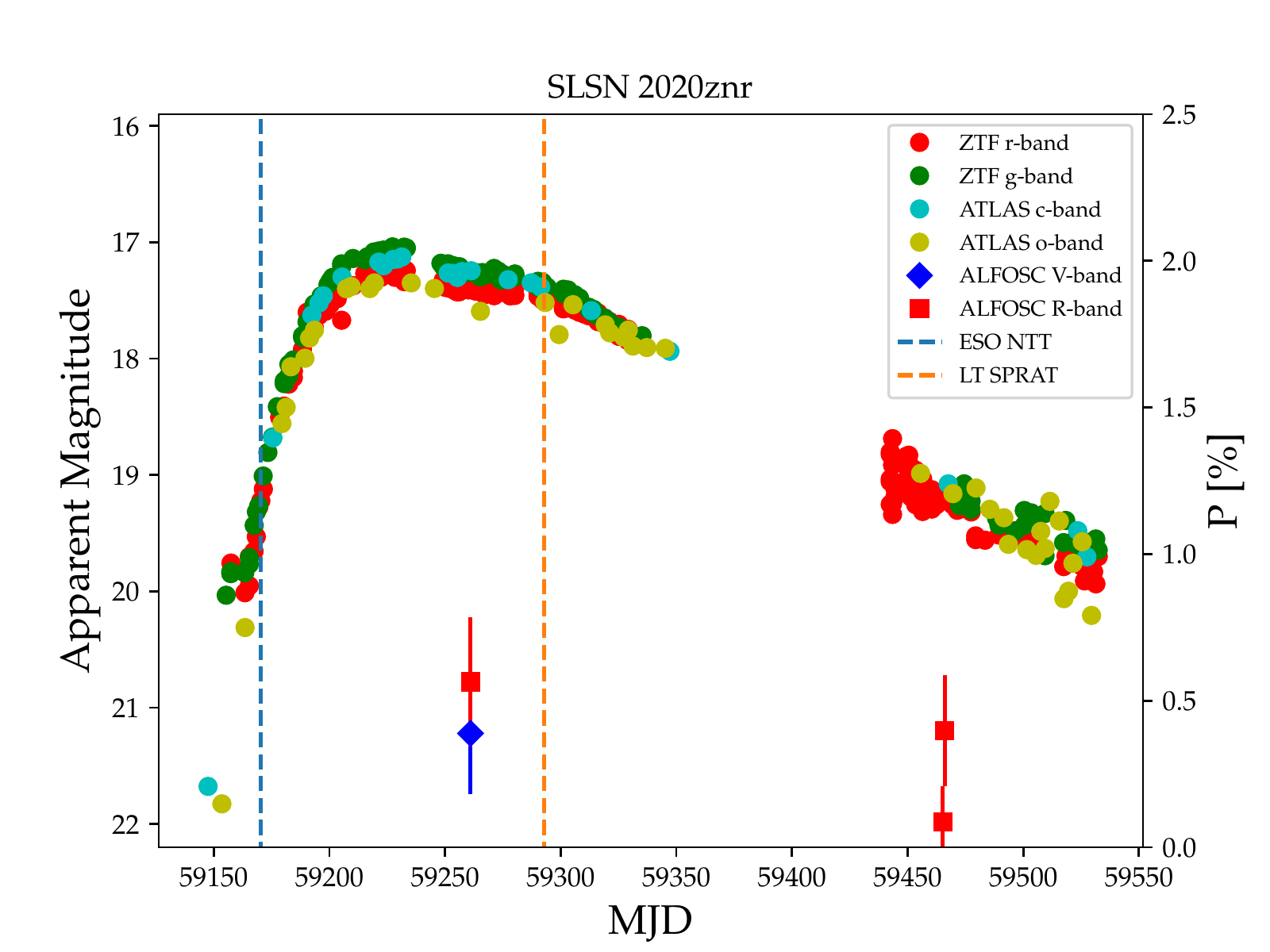}
\vspace*{0.7cm}
\caption{ZTF g- and r-band, and ATLAS c- and o-band light curve of SN 2020znr. The epochs when spectroscopy was obtained are shown with dashed-lines. The linear polarization degree, $P$, obtained at 3 epochs is also shown by the diamond and square symbols. The scale is given on the right-axis.}
\label{fig:2020znr_ZTF20acphdcg_lasair_LCs}
\end{center}
\end{figure}

The ZTF g- and r-band public photometry data obtained on SN 2020znr  
(ZTF object ZTF20acphdcg, Pan-STARRS 1 object PS20lkc, ATLAS object ATLAS20bgae, Gaia object Gaia20fkx) were retrieved from the Lasair broker \footnote{{\tt https://lasair.roe.ac.uk/object/ZTF20acphdcg/}} \citep[][]{smith2019}. The ATLAS forced photometry data obtained on ATLAS object ATLAS20bgae were retrieved from the ATLAS public server \footnote{{\tt
    https://fallingstar-data.com/forcedphot/}} \citep[][]{tonry2018}.
    They were clipped and binned to one day using the publicly available code plot$\_$atlas$\_$fp.py \footnote{{\tt
    https://gist.github.com/thespacedoctor/
    86777fa5a9567b7939e8d84fd8cf6a76}.}
    The public ZTF data and the public stacked and binned ATLAS data are displayed in Figure~\ref{fig:2020znr_ZTF20acphdcg_lasair_LCs}. They were obtained at Modified Julian Date (MJD) ranging in MJD $=[59145.6 - 59532.5]$ days. Maximum light occurred close to MJD$\approx 59227$ days as estimated from the g-band light curve. This means that polarimetry was obtained at phases $\approx$ +34, $\approx$ +238 and $\approx$ +239 days, while spectroscopy would have been obtained at phases $\approx$ -57 and $\approx$ +66 days. 
Also shown in the Figure is the linear polarization degree obtained at
one epoch in the V- and R-filters and at two close epochs in the R-filter. These results will be discussed further. A compilation of the photometry used in this work is given in the tables displayed in section~\ref{photometry_table}.

\subsection{Polarimetry} \label{data_pol}

\begin{figure}
\begin{center}
\vspace*{2mm}
\centering
\hspace*{0.cm}
\includegraphics[width=80mm,angle=0]{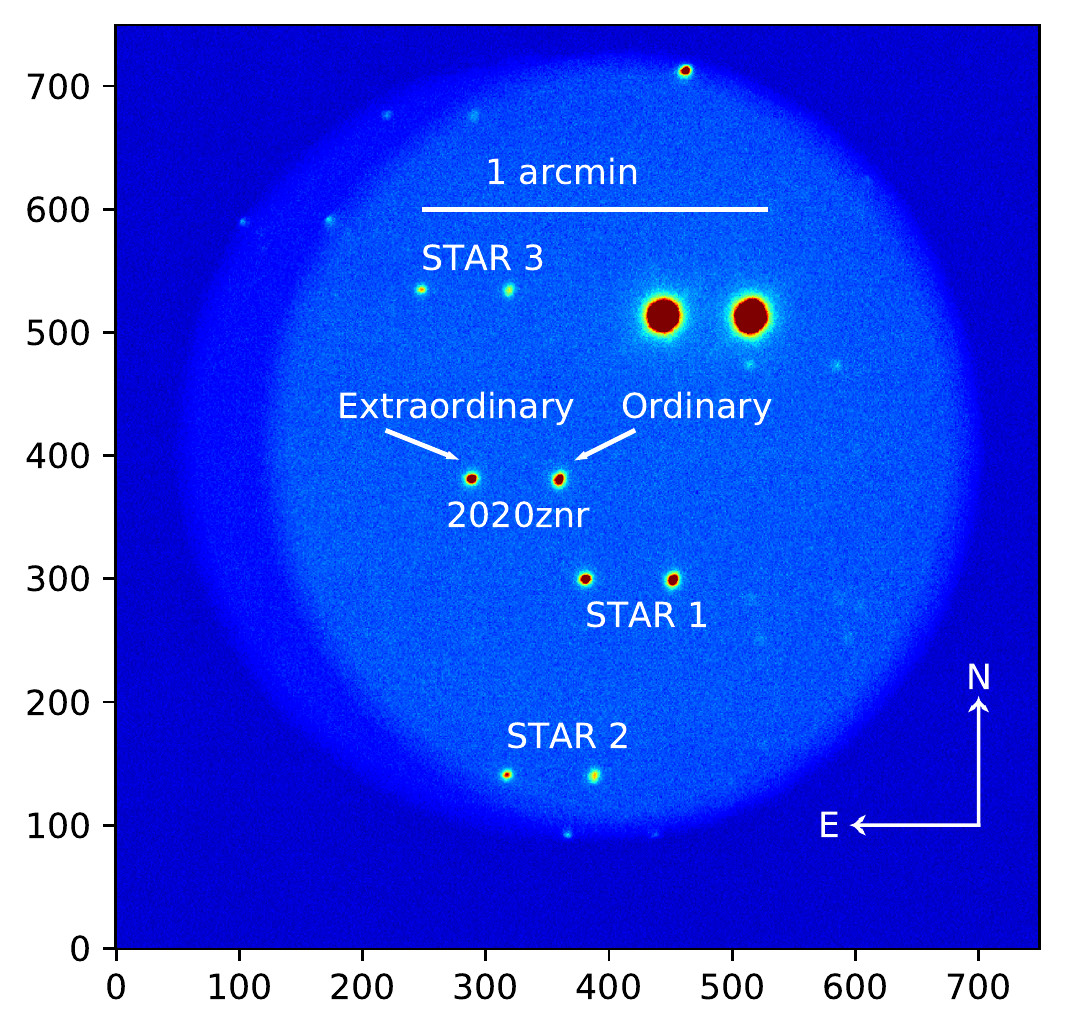}
\vspace*{0.7cm}
\caption{Linear polarimetry of SN 2020znr with ALFOSC in one of the
  V-band flat-field and bias corrected raw data frame (file ALEb150089.fits) obtained with the half-wave plate at a position angle of $0.0^{\circ}$. Each pixel embeds the number of counts obtained after an exposure of 60 seconds. Imaging polarimetry was acquired through half-wave
plates positions angles at $0.0^{\circ}$, $22.5^{\circ}$,
$45.0^{\circ}$ and $67.5^{\circ}$.  The calcite plate splits the light
from the several objects into Ordinary images and
Extraordinary images separated by about $15^{\arcsec}$ from each other.}
\label{fig:sn2020znr_image}
\end{center}
\end{figure}

The observations log of the imaging polarimetry obtained with ALFOSC on the Nordic Optical Telescope (NOT) is displayed in Table~\ref{tab:log}. Linear polarimetry
is made using a half wave plate in the FAPOL unit and a calcite plate
mounted in the aperture wheel. The calcite plate provides the
simultaneous measurement of the ordinary and the extraordinary
components of two orthogonal polarized beams (see
Figure~\ref{fig:sn2020znr_image}). The half wave
plate can be rotated in steps of 22.5$^{\circ}$ from 0$^{\circ}$ to
$337.5^{\circ}$. As a standard, 4 angles are used ($0^{\circ},
22.5^{\circ}, 45^{\circ}$, and $67.5^{\circ}$), which we used during
our observations as referred to with the factor 4 used in the
exposure time calculations displayed in Table~\ref{tab:log}.

\begin{table}
	\centering
	\caption{Observations log of the imaging polarimetry observations.}
	\label{tab:log}
	\begin{tabular}{lllcc} 
          \hline
          UT Time & Object & Exp. Time [s] & Filter & Seeing [$\arcsec$] \\
          \hline
          2021-02-16 01:07 &  BD+52913     & $ 4 \times 3 $ & V &1.7\\
          2021-02-16 01:08 &  BD+52913     & $ 4 \times 3 $ & R &1.7\\
          2021-02-16 01:11 &  HD251204     & $ 4 \times 3 $ & V &1.3\\
          2021-02-16 01:12 &  HD251204     & $ 4 \times 3 $ & R &1.3\\
          2021-02-16 01:14 &  2020znr & $ 3 \times (4 \times 60)$ & V &1.2\\
          2021-02-16 01:24 &  2020znr & $ 3 \times (4 \times 60)$ & R &1.0\\
          \hline
          2021-09-09 04:40 &  BD+52913     & $ 4 \times (4 \times 10)$ & R &0.4\\
          2021-09-09 04:47 &  HD251204     & $ 6 \times (4 \times 2)$ & R &0.6\\
          2021-09-09 05:02 &  2020znr & $ 4 \times (4 \times 180)$ & R &0.4\\
          \hline
          2021-09-10 05:29 &  BD+52913     & $ 2 \times (4 \times 10 )$ & R &0.6\\
          2021-09-10 05:34 &  HD251204     & $ 3 \times (4 \times 2)$ & R &0.6\\
          2021-09-10 05:02 &  2020znr & $ 2 \times (4 \times 180)$ & R &0.6\\
          \hline
	\end{tabular}
\end{table}

The reduction of the polarimetry data was done using our own pipeline.
Flat and bias frames obtained during the nights of the observations were cropped at the size of the observation frames. The observation frames were then bias subtracted and flat-fielded using median frames of the bias and flat images, respectively.
The zero level of each reduced observation frame was assessed using
the python package PHOTUTILS then DAOfind was used to produce a list
of the sources detected above a given threshold in each final frame. 
Circular aperture photometry was then operated on each detected source
taking into account the average seeing information. 
Aperture photometry is preferred over PSF fitting because the polarized images are not circular after the half wave plate. These apertures cover up to about 2 to 3 times the full width at half maximum (FWHM). At this stage a total flux in Analog-to-Digital Units (ADU) was obtained on each ordinary and
extraordinary images of the astrophysical sources of interest.

Raw Stokes $Q$ and $U$ parameters were calculated using the formalism
provided by \citet{magalhaes1984} and \citet{ramirez2017}. From each
reduced frame obtained at a half waveplate angle position, $i$, we
first calculate the ordinary and extraordinary fluxes, $F_{o,i}$ and
$F_{e,i}$, respectively. The modulation of the intensity with the
half waveplate position, $z_{i}$, is given by the expression:
\begin{equation}
z_{i}= \frac {F_{e,i}-F_{o,i} (F_{e}^{T}/F_{o}^{T}) }{F_{e,i} + F_{o,i}
 (F_{e}^{T}/F_{o}^{T}) } = Q cos(4 \psi_{i}) + U sin(4 \psi_{i}),   
\end{equation}
where, $F_{o}^{T}$, and , $F_{e}^{T}$, are the total ordinary and
extraordinary fluxes, respectively, summed over the waveplate
positions and, $\psi_{i}$, is the half-wave plate position angle
at position, $i$.
Following the prescription given by \citet{magalhaes1984}
the solution for the Stokes parameters $Q$ and $U$ are:
\begin{equation}
  Q = \frac{2}{\mu} \sum_{i}^{\mu}z_{i}cos(4 \psi_{i}),
\end{equation}
\begin{equation}
  U = \frac{2}{\mu} \sum_{i}^{\mu}z_{i}sin(4 \psi_{i}),
\end{equation}
where, $\mu$, is the total number of half-wave plate positions.
Once the Stokes parameters are calculated, one can directly obtain
the fraction of polarization:
\begin{equation}
  P= \sqrt{Q^2+U^2} ,
\end{equation}
and the polarization angle position:
\begin{equation}
 \theta = \frac{1}{2} tan^{-1}\frac{U}{Q},
\end{equation}
where $\theta$ is counted positively from north to east in the
equatorial reference frame after correction of the zero-angle
calibrated with a polarized standard star.
Assuming the uncertainties on Stokes parameters, $Q$, and, $U$,
are of the same order and therefore that
$\sigma_{Q}$ = $\sigma_{U}$ = $\sigma_{P}$,   
we follow \citet{magalhaes1984} and \citet{naghizadeh-khouei1993} and
calculate the errors as:
\begin{equation}
\sigma_{P} = \frac{1}{\sqrt{\mu-2}} \sqrt{\frac{2}{\mu}\sum_{i}^{\mu} z_{i}^{2} - Q^{2} - U^{2}},
\end{equation}
\begin{equation}
\sigma_{\theta} = 28^{\circ}.65 \frac{\sigma_{P}}{P}.
\end{equation}

In the following we will discuss the fraction of polarization obtained
without applying any debiasing method. A test of our polarization data reduction method is given in section~\ref{pipeline} where the output of our pipeline is compared to the results obtained by \citet{lee2020} on SN 2020ank ALFOSC polarimetry data. Field stars, STAR 1, STAR 2 and STAR 3, displayed in Figure~\ref{fig:sn2020znr_image} are field stars that will be considered to make estimates of the Milky Way Interstellar Polarization (ISP).

\section{Analysis}

\subsection{Light Curve Modelling} \label{lc_modelling}

We fit the multi-band light curve using the magnetar spin-down model in the Modular Open-Source Fitter for Transients (\texttt{MOSFiT}) code, which uses a Markov Chain Monte Carlo (MCMC) algorithm to perform Bayesian parameter estimation for supernova light curves \citep{2018ApJS..236....6G}.  We use the Dynesty sampler \citep{2020MNRAS.493.3132S, 2019S&C....29..891H}, which utilizes dynamic nested sampling.  The uncertainty presented is only the statistical uncertainty in the fits, and does not include systematic uncertainty inherent in the simplified one-zone \texttt{MOSFiT} model.

\begin{figure*}
\begin{center}
\centering
\includegraphics[width=0.9\textwidth,angle=0]{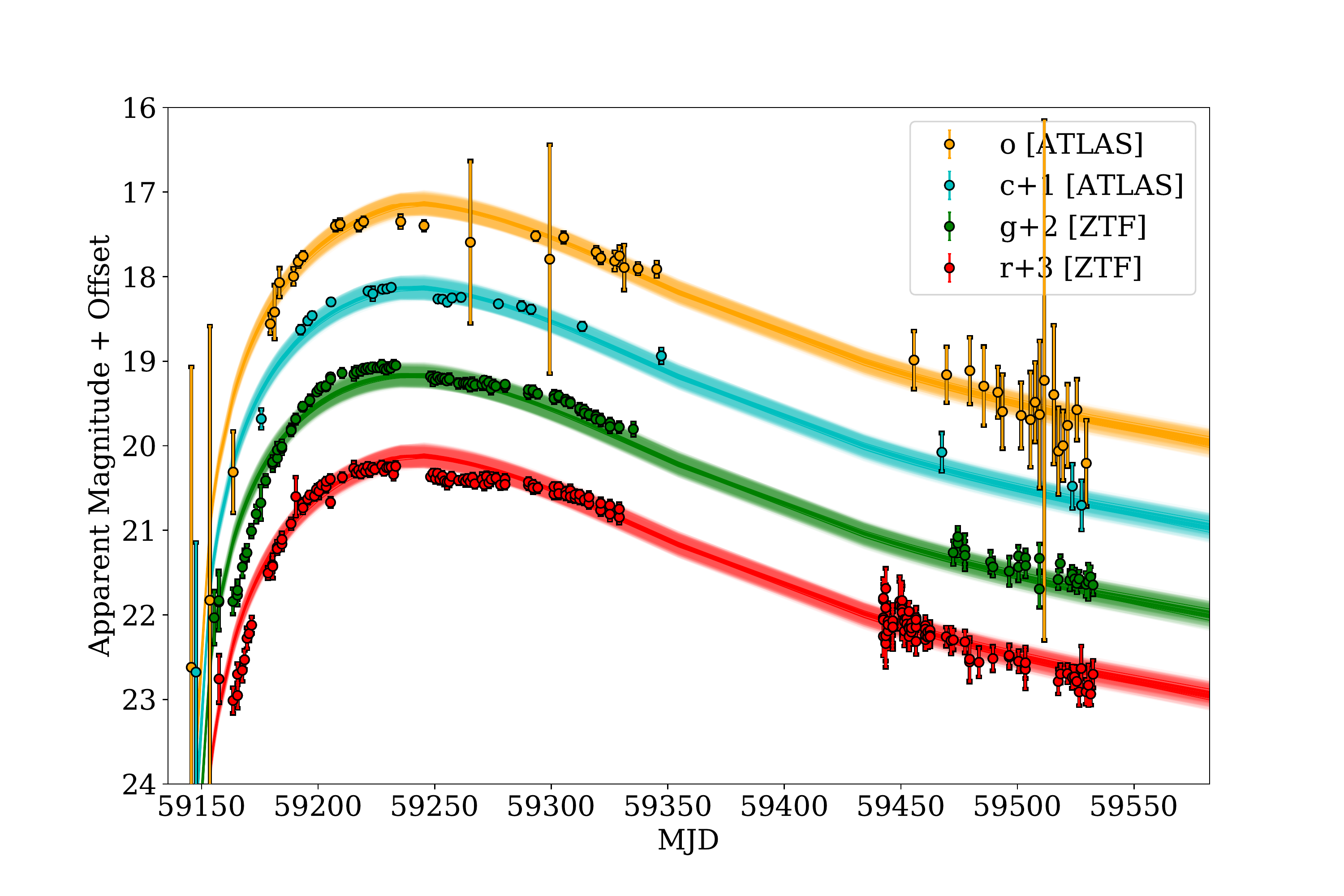}
\caption{Multi-band light curve of SN 2020znr inferred from the magnetar-model, with each band offset for clarity.  The filled area shows the range of most likely models generated by \texttt{MOSFiT}. See Section \ref{lc_modelling} for details.}
\label{fig:mosfitfull}
\end{center}
\end{figure*}

\begin{table*}
	\centering
	\caption{Median and 1$\sigma$ best fit parameters for magnetar models. Parameters obtained on SN 2020znr are discussed in Section \ref{lc_modelling} and compared to those obtained on SN 2015bn and SN 2017egm in Section \ref{discussion}.}
	\label{tab:mosfitparams}
	\begin{tabular}{ccccccc} 
          Parameter Symbol & Definition & Prior & Best Fit Value & Best Fit Value & Best Fit Values & Units\\
           &  & 2020znr & 2020znr & 2015bn & 2017egm & \\
           &  & This work & This work & \citet{2017ApJ...850...55N} & \citet{2017ApJ...845L...8N} & \\
          \hline
          $B_\perp$ & Magnetar Magnetic Field Strength & [0.1,10] & $0.51^{+0.10}_{-0.13}$ &$0.31^{+0.07}_{-0.05}$& [0.7 – 1.7] & 10$^{14}$ G\\
          $M_{\rm NS}$ & Neutron Star Mass & [1.0,2.0] & $1.68^{+0.21}_{-0.31}$ &$1.78^{+0.28}_{-0.23}$& ... & $M_\odot$\\
          $P_{\rm spin}$ & Magnetar Spin Period & [1,10] & $2.80^{+0.26}_{-0.39}$ &$2.16^{+0.29}_{-0.17}$& [4–6]& ms \\
          $\log (M_{\rm ej})$ & Ejecta Mass & [-1,2] & $1.33^{+0.03}_{-0.03}$ &$1.09^{+0.08}_{-0.13}$& [0.3–0.6] & $M_\odot$\\
          $v_{\rm ej}$ &  Ejecta Velocity & [1,20] & $5.56^{+0.13}_{-0.13}$ &$5.46^{+0.16}_{-0.14}$&...& 10$^3$ km/s\\
          \hline
	\end{tabular}
\end{table*}

\begin{figure*}
\begin{center}
\centering
\includegraphics[width=0.8\textwidth,angle=0]{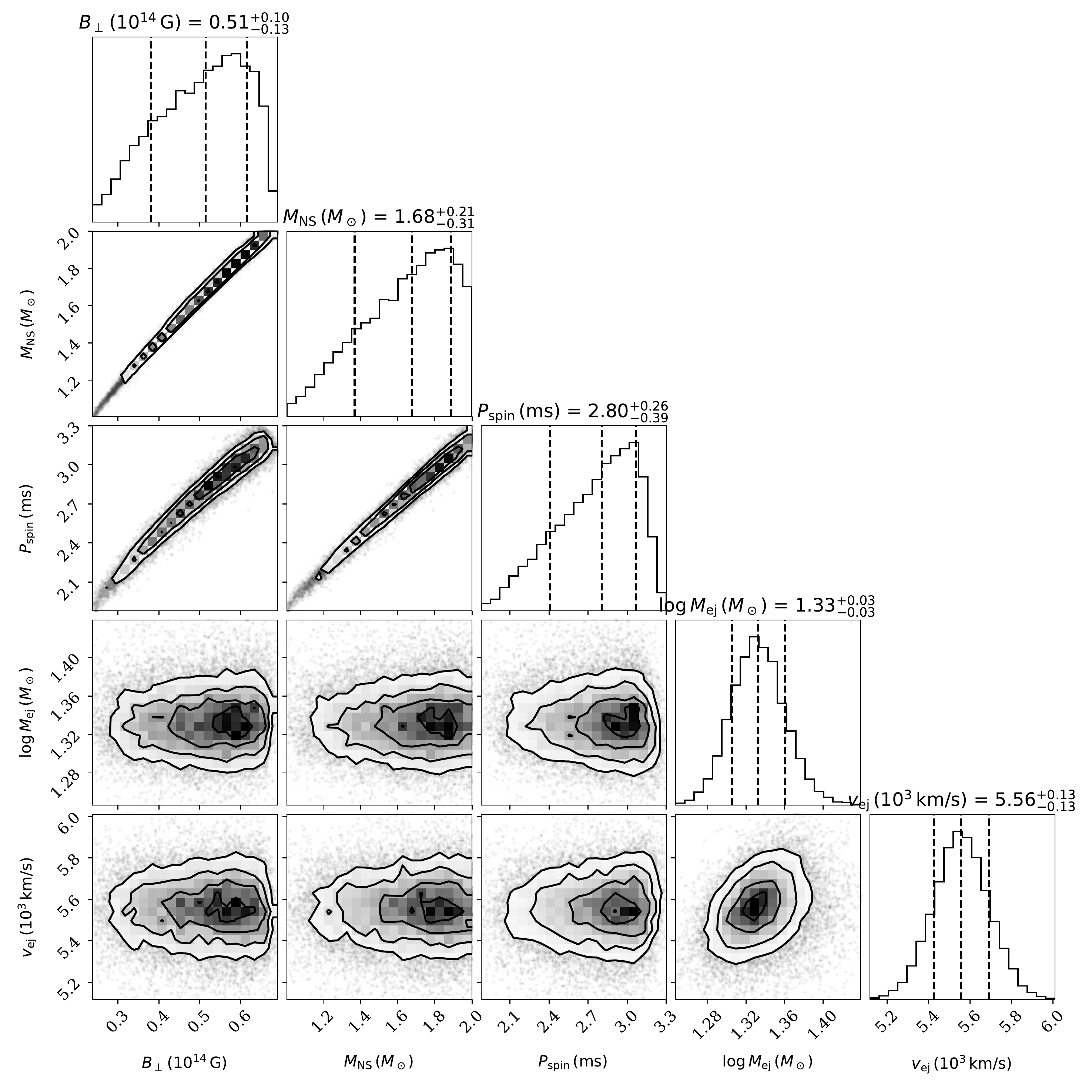}
\caption{1D and 2D posterior distributions of the
magnetar model parameters. Median and 1$\sigma$ values are
marked and labeled - these are used as the best fit values (discussed in text).} 
\label{fig:mosfitpost}
\end{center}
\end{figure*}

The magnetar-model fits are shown in Figure \ref{fig:mosfitfull} and the most physically relevant parameters are listed in Table \ref{tab:mosfitparams}, with their posteriors shown in Figure \ref{fig:mosfitpost}.  
The model provides a good fit to the data, although it does overestimate luminosity during the pre-peak rise. The physical parameters we find are $B_\perp \approx 5 \times 10^{13}$ G, M$_{\rm NS} \approx 1.7 M_\odot$, $P_{\rm spin} \approx 3.0$ ms, $M_{\rm ej} \approx 21 M_\odot$, and $v_{\rm ej} \approx 5500$ km/s, where $B_\perp$ is the magnetar magnetic field strength, M$_{\rm NS}$ is the neutron star mass, $P_{\rm spin}$, is magnetar spin period, $M_{\rm ej}$ is the ejecta mass and, $v_{\rm ej}$ is the ejecta velocity. 
These best fit parameters and uncertainties are the median and 1$\sigma$ values from the one-dimensional posterior for each of the parameters.  While the median value is close to the most-likely value for $M_{\rm ej}$ and $v_{\rm ej}$, which have symmetric posteriors, the median value is lower than the most-likely value by $\sim$ 1$\sigma$ for $B_\perp$, M$_{\rm NS}$, and $P_{\rm spin}$, which are asymmetric around the median value (this is also seen in the joint posterior for all SLSNe examined by \citet{2017ApJ...850...55N}) and feature a steady increase in probability as the parameters increase, followed by a sharp cutoff.  This cutoff is likely due to M$_{\rm NS}$ reaching the maximum of its prior.
These parameters are fairly typical of SLSNe-I \citep{2017ApJ...850...55N, 2018ApJ...869..166V, 2020ApJ...897..114B, 2021arXiv210906970Y}, except for the ejecta mass, which is larger than average, although not abnormally so, as will be discussed further in Section \ref{discussion}.  However, SN 2020znr has a broad light curve and fades slowly, which is indicative of a large ejecta mass.
The total kinetic energy of the ejecta is calculated to be $E_{\rm K} \approx 4 \times 10^{51}$ ergs.
The mass of the progenitor star, $M_* = M_{\rm NS} + M_{\rm ej} \approx 23 M_\odot$, which is consistent with the 3.6-40 $M_\odot$ range inferred in the mass distribution found by \cite{2020ApJ...897..114B}.

We find strong correlations between $B_\perp$, $M_{\rm NS}$, and $P_{\rm spin}$ in the 2D posterior distributions of these parameters (see Figure \ref{fig:mosfitpost}).  These correlations have been observed in the posteriors of magnetar model fits of some supernovae \citep[e.g., SN2015bn;][]{2017ApJ...850...55N}, but are not seen in others \citep[e.g., SN2018lfe;][]{2021arXiv210906970Y}. It is unknown why only some SLSNe show these correlations, and while this bimodality could have a physical origin, it is likely due some supernovae being more well sampled and having stronger constraints on system parameters. 



\subsection{Polarimetry null-detection} \label{pol_analysis}

\begin{table*}
	\centering
	\caption{Polarimetry results. $^{\rm(a)}$: Stokes
          parameters, $\overline{Q}$ and $\overline{U}$, directly obtained
          from the ALFOSC data frames Extraordinary and Ordinary images 
          without applying any further corrections. $^{\rm(b)}$: instrumental
          polarization estimates. $^{\rm(c)}$: instrumental
          polarization corrected. Assuming HD251204 has a polarization
          angle of $147 ^{\circ}$ in V- and R-bands the zero
          polarization angle, $ZPA$, estimates are $83.9 ^{\circ}$ and
          $87.7 ^{\circ}$, respectively. $^{\rm(d)}$: IP and $ZPA$
          corrected. $^{\rm(e)}$: $IP$ and $ZPA$ corrected +
          Milky Way Interstellar Polarization
          corrected with Stokes parameters obtained on star 2$^{\rm(f)}$ (see values shown in bold and underlined, and see text for details), 
          which is at the largest heliocentric distance, then with weighted average Stokes parameters obtained on stars STAR 1, STAR 2 and STAR 3 $^{\rm(g)}$ (See values only shown in bold and see text for details).}
	\label{tab:results}
	\begin{tabular}{lccccccccccc} 
          \hline
          Date
          &Source
          &filter
          & $\overline{Q}^{\rm (a)}$
          & $\overline{U}^{\rm(a)}$
          & $P [\%]^{\rm(b)} $
          & $P [\%]^{\rm(c)}$
          & $\theta [^{\circ}]^{\rm(c)}$
          & $P [\%]^{\rm(d)} $
          & $\theta [^{\circ}]^{\rm(d)}$
          & $P [\%]^{\rm(e)} $\\
          \hline
2021-02-16 & BD$+$52913 & V & -0.08 & 0.07 & \textbf{ 0.11 $\pm$ 0.07 }& ... & ... & ... & ... & ...  \\
 ... & HD251204 & V & -2.91 & 3.93 & ... & \textbf{ 4.78 $\pm$ 0.13 }& \textbf{ 63.13 $\pm$ 0.75 }& ... & ... & ... \\
 ... & STAR 1 & V & -0.35 & 0.22 & ... &....&...& 0.31 $\pm$ 0.17 & 159.23 $\pm$ 15.56  & ...  \\
 ... & STAR 2 & V & 0.14 & 0.33 & ... &....&...& \textbf{\underline{0.34 $\pm$ 0.28}}  & \textbf{\underline{108.06 $\pm$ 24.11}}  & ...  \\
 ... & STAR 3 & V & 0.53 & -0.00 & ... &....&...& 0.62 $\pm$ 0.22 & 80.42 $\pm$ 10.35  & ...  \\
 ... & ISP & V & 0.00 & 0.00 & ... &....&...& \textbf{ 0.06 $\pm$ 0.13 }& \textbf{ 102.38 $\pm$ 66.00 }& ...  \\
 ... & 2020znr $^{\rm(f)}$& V & -0.42 & -0.25 & ... &....&...& ...&...& \textbf{ \underline{0.81 $\pm$ 0.34}  } \\
 ... & 2020znr $^{\rm(g)}$& V & -0.42 & -0.25 & ... &....&...& ...&...& \textbf{ 0.44 $\pm$ 0.21 } \\
2021-02-16 & BD$+$52913 & R & -0.06 & -0.10 & \textbf{ 0.12 $\pm$ 0.16 }& ... & ... & ... & ... & ...  \\
 ... & HD251204 & R & -2.49 & 4.29 & ... & \textbf{ 5.01 $\pm$ 0.24 }& \textbf{ 59.50 $\pm$ 1.37 }& ... & ... & ... \\
 ... & STAR 1 & R & -0.29 & -0.04 & ... &....&...& 0.23 $\pm$ 0.19 & 169.75 $\pm$ 24.07  & ...  \\
 ... & STAR 2 & R & -0.09 & -0.23 & ... &....&...& \textbf{\underline{0.14 $\pm$ 0.22}} & \textbf{\underline{36.47 $\pm$ 44.48}}  & ...  \\
 ... & STAR 3 & R & -0.41 & 0.21 & ... &....&...& 0.46 $\pm$ 0.32 & 156.43 $\pm$ 19.75  & ...  \\
 ... & ISP & R & 0.00 & 0.00 & ... &....&...& \textbf{ 0.06 $\pm$ 0.14 }& \textbf{ 169.67 $\pm$ 65.05 }& ...  \\
 ... & 2020znr $^{\rm(f)}$& R & -0.51 & -0.17 & ... &....&...& ...&...& \textbf{ \underline{0.43 $\pm$ 0.32 }} \\
 ... & 2020znr $^{\rm(g)}$& R & -0.51 & -0.17 & ... &....&...& ...&...& \textbf{ 0.59 $\pm$ 0.22 } \\
2021-09-08 & BD$+$52913 & R & -0.05 & 0.00 & \textbf{ 0.05 $\pm$ 0.03 }& ... & ... & ... & ... & ...  \\
 ... & HD251204 & R & -2.51 & 4.18 & ... & \textbf{ 4.85 $\pm$ 0.04 }& \textbf{ 60.26 $\pm$ 0.26 }& ... & ... & ... \\
 ... & STAR 1 & R & -0.25 & 0.08 & ... &....&...& 0.21 $\pm$ 0.07 & 166.57 $\pm$ 9.61  & ...  \\
 ... & STAR 2 & R & -0.09 & -0.00 & ... &....&...& \textbf{\underline{0.04 $\pm$ 0.11}} & \textbf{\underline{0.03 $\pm$ 73.54}}  & ...  \\
 ... & STAR 3 & R & -0.24 & -0.08 & ... &....&...& 0.20 $\pm$ 0.11 & 7.91 $\pm$ 16.17  & ...  \\
 ... & ISP & R & 0.00 & 0.00 & ... &....&...& \textbf{ 0.05 $\pm$ 0.06 }& \textbf{ 174.86 $\pm$ 32.60 }& ...  \\
 ... & 2020znr $^{\rm(f)}$& R & 0.12 & -0.06 & ... &....&...& ...&...& \textbf{ \underline{0.22 $\pm$ 0.14 }} \\
 ... & 2020znr $^{\rm(g)}$& R & 0.12 & -0.06 & ... &....&...& ...&...& \textbf{ 0.09 $\pm$ 0.11 } \\
2021-09-09 & BD$+$52913 & R & -0.05 & 0.13 & \textbf{ 0.14 $\pm$ 0.07 }& ... & ... & ... & ... & ...  \\
 ... & HD251204 & R & -2.48 & 4.24 & ... & \textbf{ 4.78 $\pm$ 0.08 }& \textbf{ 60.31 $\pm$ 0.48 }& ... & ... & ... \\
 ... & STAR 1 & R & -0.22 & -0.02 & ... &....&...& 0.23 $\pm$ 0.13 & 16.63 $\pm$ 16.51  & ...  \\
 ... & STAR 2 & R & -0.42 & -0.30 & ... &....&...& \textbf{\underline{0.56 $\pm$ 0.09}} & \textbf{\underline{21.15 $\pm$ 4.64}}  & ...  \\
 ... & STAR 3 & R & -0.11 & -0.12 & ... &....&...& 0.26 $\pm$ 0.13 & 34.36 $\pm$ 14.56  & ...  \\
 ... & ISP & R & 0.00 & 0.00 & ... &....&...& \textbf{ 0.12 $\pm$ 0.07 }& \textbf{ 22.94 $\pm$ 16.04 }& ...  \\
 ... & 2020znr $^{\rm(f)}$& R & -0.05 & -0.29 & ... &....&...& ...&...& \textbf{ \underline{0.37 $\pm$ 0.20 }} \\
 ... & 2020znr $^{\rm(g)}$& R & -0.05 & -0.29 & ... &....&...& ...&...& \textbf{ 0.40 $\pm$ 0.18 } \\
          \hline
	\end{tabular}
\end{table*}

Using the formalism described in section~\ref{data_pol} we first
calculated the average raw Stokes parameters, $\overline{Q}^{\rm (a)}$, and
$\overline{U}^{\rm (a)}$, obtained with Bessel -V or -R filters
on the polarized and unpolarized stars, on 
the three field stars identified in
Figure~\ref{fig:sn2020znr_image}, and on the main target SN 2020znr.
Their values are displayed in columns 4 and 5 in
Table~\ref{tab:results}. The Instrumental Polarization ($IP$) of the
ALFOSC imaging polarimeter estimated with the unpolarized star,
BD$+52913$, shows Stokes $Q$ and $U$ of order 0.1$\%$
or below (see columns 6  in Table~\ref{tab:results}).
These values were subtracted from the raw Stokes
parameter estimates on all other targets. 
The calibration of the Zero
Polarization Angle ($ZPA$) of the experiment was carried out using the
polarized star HD251204 (see columns 7 and 8 in
Table~\ref{tab:results}) . Assuming HD251204 has a polarization
angle of $147 ^{\circ}$ \citep[][]{turnshek1990} in both V- and R-bands we got , $ZPA$, estimates of
$83.9 ^{\circ}$ and $87.7 ^{\circ}$, respectively. These values
are subsequently used to correct for the IP corrected Stokes $Q$ and $U$ of the remaining sources of
interest (STAR 1, 2 and 3) 
as shown in columns 9 and 10 in Table~\ref{tab:results}. 

To obtain the level of linear polarization of
SN 2020znr, the remaining step is to assess the level of polarization
produced by the interstellar medium in our galaxy. To do so we first cross-matched the coordinates of fields stars 1, 2 and 3 with the GAIA Early Data Release 3 (EDR3) distances catalog \citep[][]{bailer-jones2021}. The median of the geometric distance posterior, $rgeo$, and the median of the photogeometric distance posterior, $rpgeo$, ranging from about 2000 pc to about 4000 pc are displayed in Table~\ref{tab:gaia_edr3_dist}.
In addition to fields stars 1, 2 and 3 observed with ALFOSC we also searched for candidates in the \citet{heiles2000} agglomeration file catalog. This catalog is a compilation of linear polarization measurements, mainly in the V-band, on relatively bright Galactic stars. Our main selection criterion was the angular distance to SN 2020znr as shown in the last column of Table~\ref{tab:heiles}. The closest star is HD 56986, at an angular distance 0.70$^{\circ}$, which we rejected from the Table since it is a spectroscopic binary. All the other stars are at an angular distance lower than 2 $^{\circ}$ from SN 2020znr at heliocentric distance between 50 pc and 800 pc. They have very low polarization degrees consistent with null-polarization within the uncertainties. Since all the detections are less than 1 $\sigma$ detections, are at heliocentric distances lower than 1000 pc and more importantly lie at angular distances higher than 1 $^{\circ}$ from SN 2020znr, we followed the recommendation given by \cite{tran1995} and didn't include them for estimating the Milky Way ISP. STAR 2 is at the higher heliocentric distance and it is therefore our best candidate for estimating our Galaxy ISP contribution. Hence, we first made estimates of SN 2020znr intrinsic polarization degree by subtracting its $IP$ and $ZPA$ corrected Stokes parameters from the $IP$ and $ZPA$ corrected Stokes parameters obtained on SN 2020znr. Since one can expect variations from one line-of-sight to the other, as a second step, the $IP$ and $ZPA$
corrected Stokes parameters obtained on field stars STAR 1, STAR 2 and STAR 3 were weighted averaged by using the uncertainties to determine the weights, to get a statistical estimate of the Milky Way ISP along line-of-sights probing column densities up to about 2000 pc to 4000 pc, and subtracted from the $IP$ and $ZPA$ corrected Stokes parameters obtained on SN 2020znr.

\begin{table}
	\centering
	\caption{Gaia EDR3 distances to the field stars, STAR 1, STAR 2 and STAR 3, displayed in Figure~\ref{fig:sn2020znr_image}. Parameter, $rgeo$, is the geometric distance, while parameter, $rpgeo$, is the photogeometric distance \citep[see][for details]{bailer-jones2021}}
	\label{tab:gaia_edr3_dist}
	\begin{tabular}{ccccccc} 
          \hline
          Star name &  RA (J2000) &  Dec (J2000)  & $rgeo$ & $rpgeo$ \\
           &   $[^{\circ}]$&  $[^{\circ}]$ & [pc] &  [pc] \\
          \hline
          STAR 1 & 109.7707 & 23.8805 & 2018.19177 & 1997.0564 \\ 
          STAR 2 & 109.7749 & 23.8710 & 4084.69873 & 3371.26392 \\ 
          STAR 3 & 109.7707 & 23.8945 & 2114.16211 & 2391.48169 \\ 
          \hline
	\end{tabular}
\end{table}

The final levels of polarization obtained on SN 2020znr are
displayed in the last column of Table~\ref{tab:results} 
and both methods lead to similar conclusions:
all the results are consistent with a null level of polarization within the uncertainties (i.e. all polarization signal-to-noise ratio are $< 3 \sigma$).
However, the variation of these constraints on the final level of
polarization reflects the various total integration times defined in our observing
strategy displayed in Table~\ref{tab:results}. Another important
factor is the seeing, which changed from one night to the
other. From these two points of view, the first observations obtained
in the R- and V-bands on February 2021 were done when
SN 2020znr apparent magnitudes were of order 17.5 in the ZTF r-band
(see Figure~\ref{fig:2020znr_ZTF20acphdcg_lasair_LCs}).
At that time we used relatively short integration times on
the unpolarized standard star, BD$+$52913, and the total integration time on SLSN 2020znr and the 3 fields stars was of 12 minutes.
The DESI Legacy Survey DR9 apparent magnitudes of STAR 1, STAR 2 and STAR 3
displayed in Figure~\ref{fig:sn2020znr_image}
are 17.41, 18.43, 18.95 in the g-band, and
16.84, 17.86, 18.12 in the r-band, respectively, i.e. at about the
same level or less bright than SN 2020znr at that time. Propagating
the instrumental polarization (IP) or order 0.1$\%$ on the $Q$, and
$U$, Stokes parameters of order 0.3$\%$ to 0.6$\%$ obtained on each
of these stars lead to relatively
high estimates of the polarization level, but once combined together,
the measurements give a weighted average level ISP of 0.06 $\pm$
0.13$\%$, in both bands.
A result which is in line with the very low level of polarization obtained from starlight polarization on brighter stars in the vicinity of SN 2020znr on a square area of size 5$^{\circ}$ centred on its coordinates (see Table~\ref{tab:heiles}). Using only STAR 2 the Milky Way ISP contribution is found to be slightly higher, of 0.34 $\pm$
0.28$\%$ in V-band, and of 0.14 $\pm$ 0.22$\%$ in R-band.

\begin{figure*}
\begin{center}
\vspace*{2mm}
\centering
\hspace*{0.cm}
\includegraphics[width=80mm,angle=0]{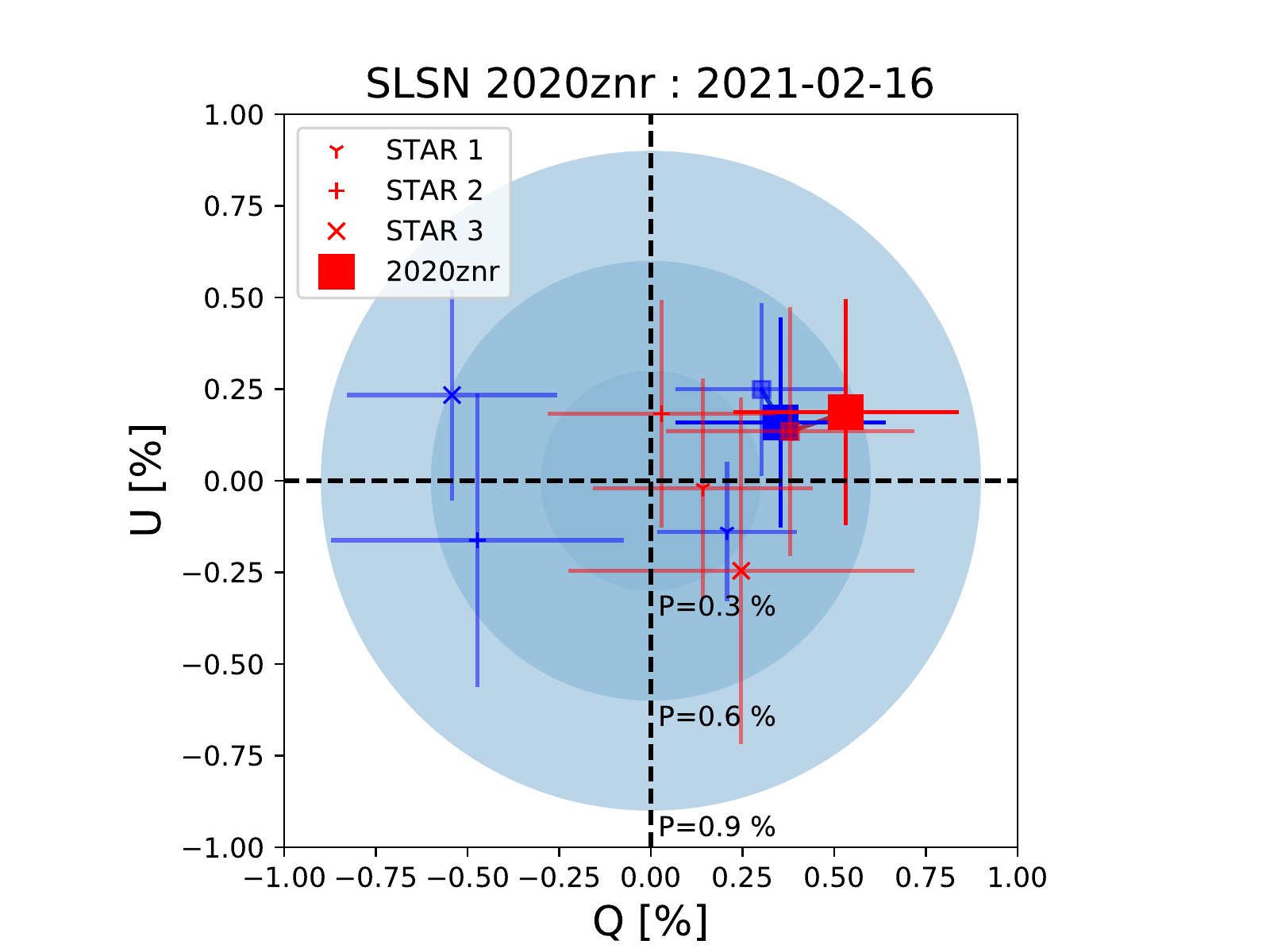}
\includegraphics[width=80mm,angle=0]{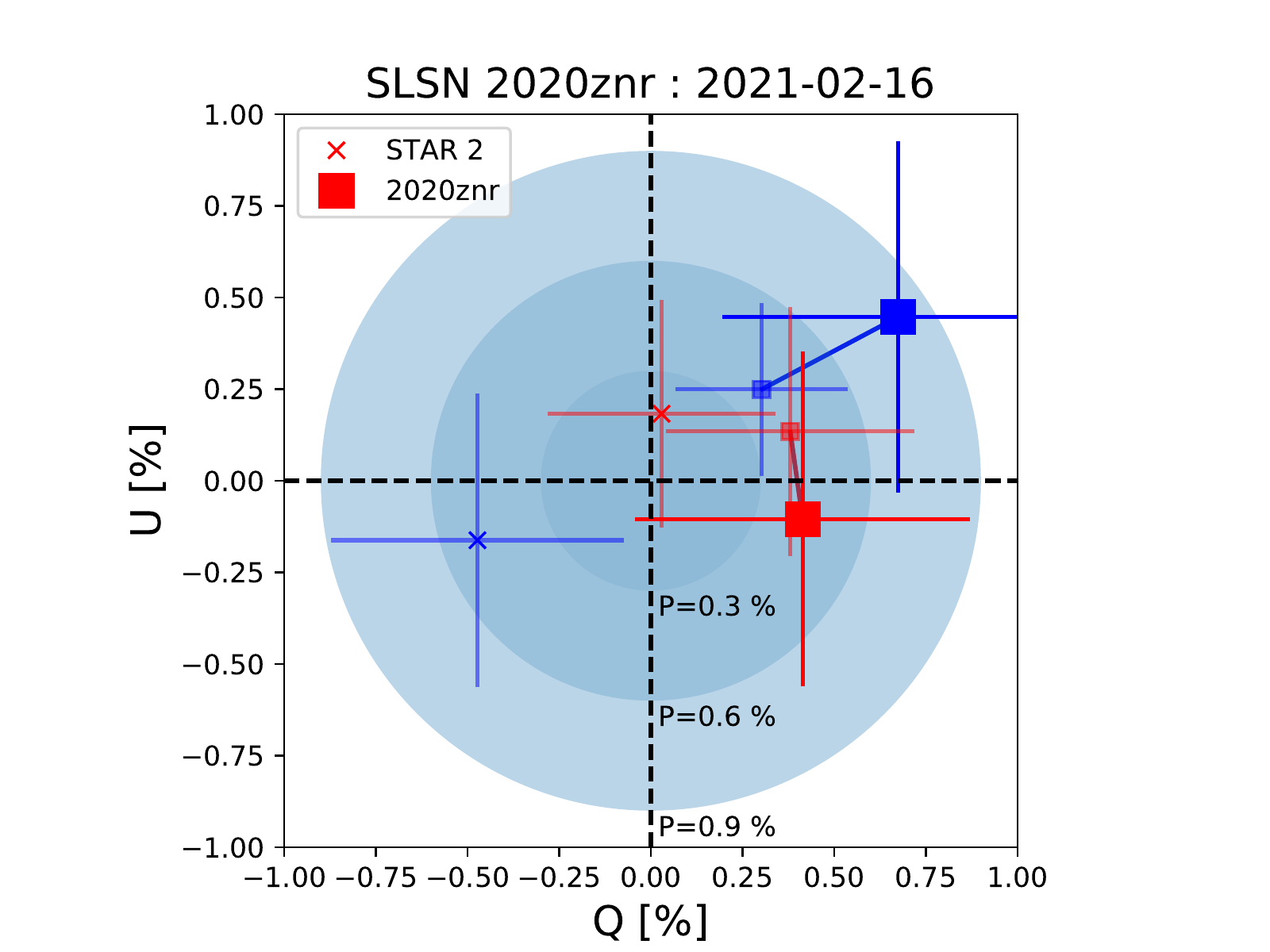}
\includegraphics[width=80mm,angle=0]{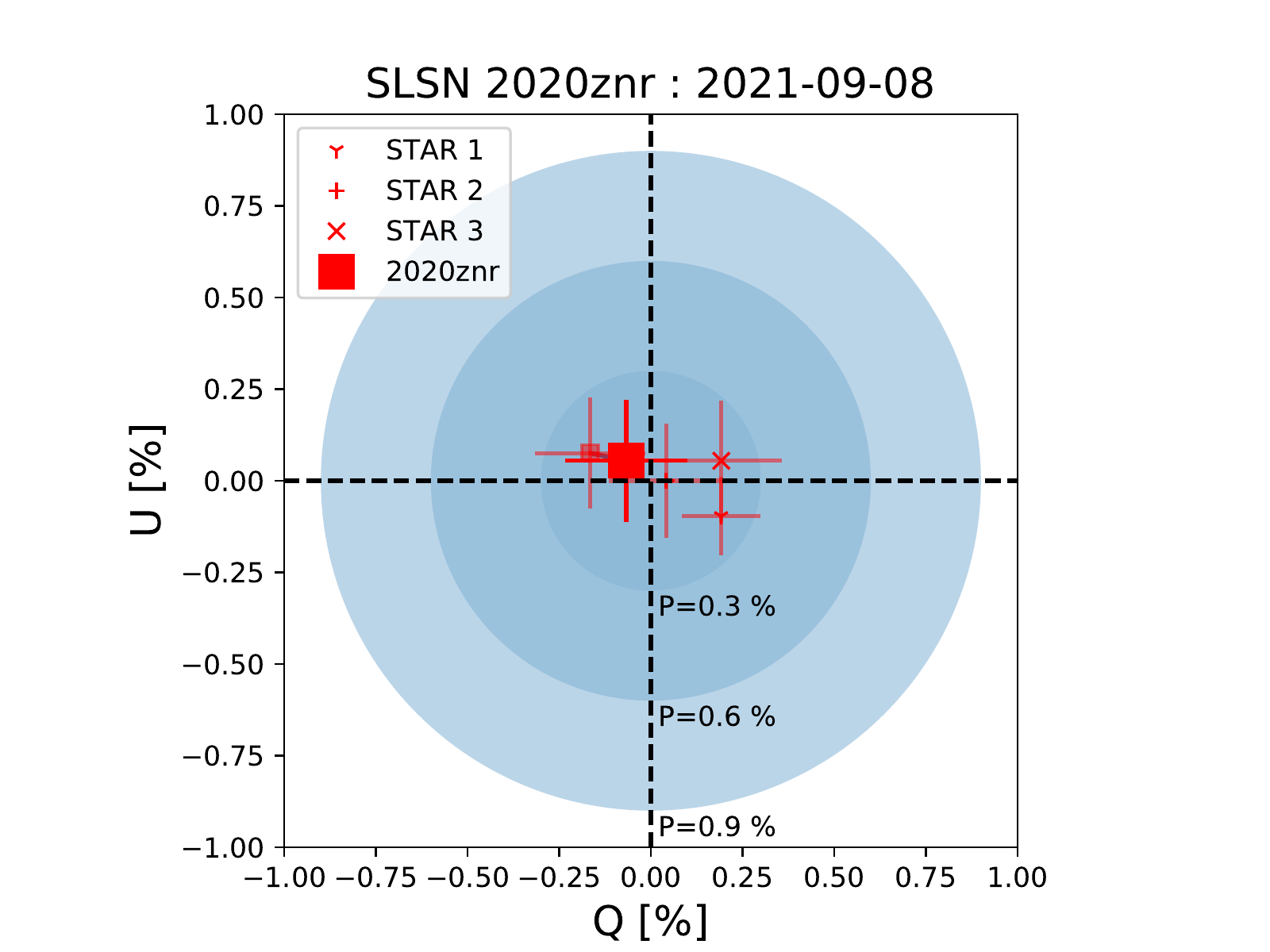}
\includegraphics[width=80mm,angle=0]{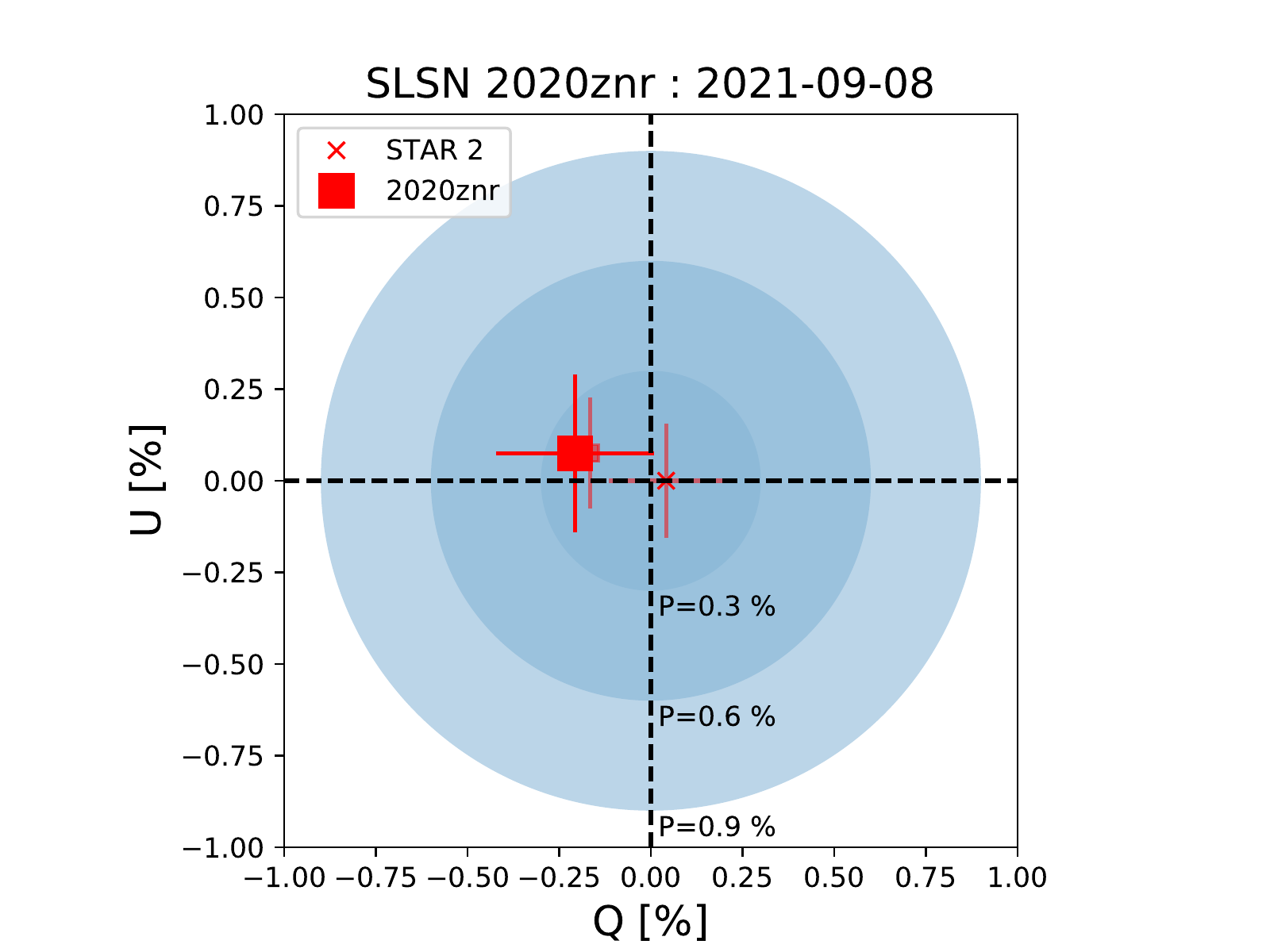}
\includegraphics[width=80mm,angle=0]{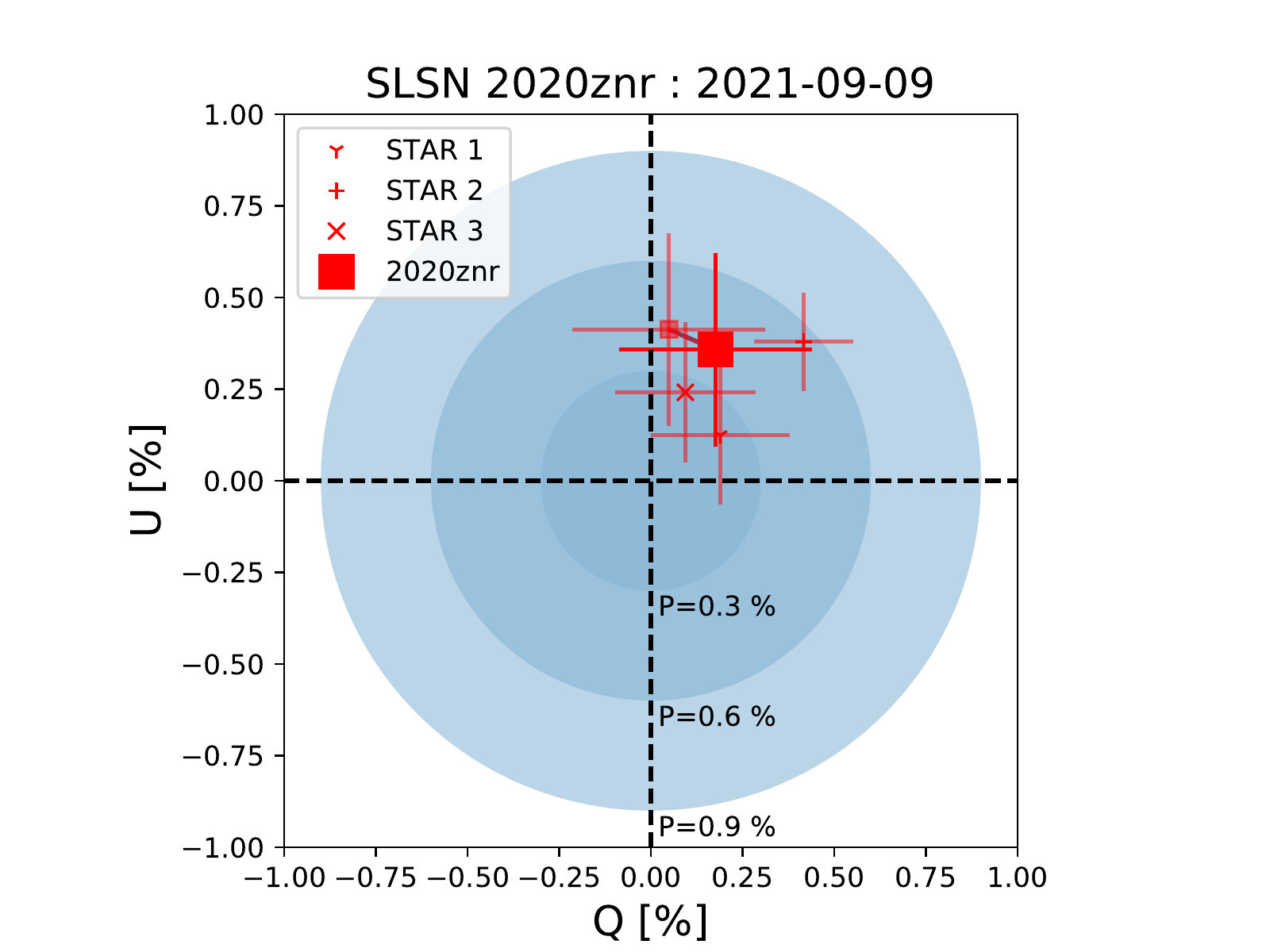}
\includegraphics[width=80mm,angle=0]{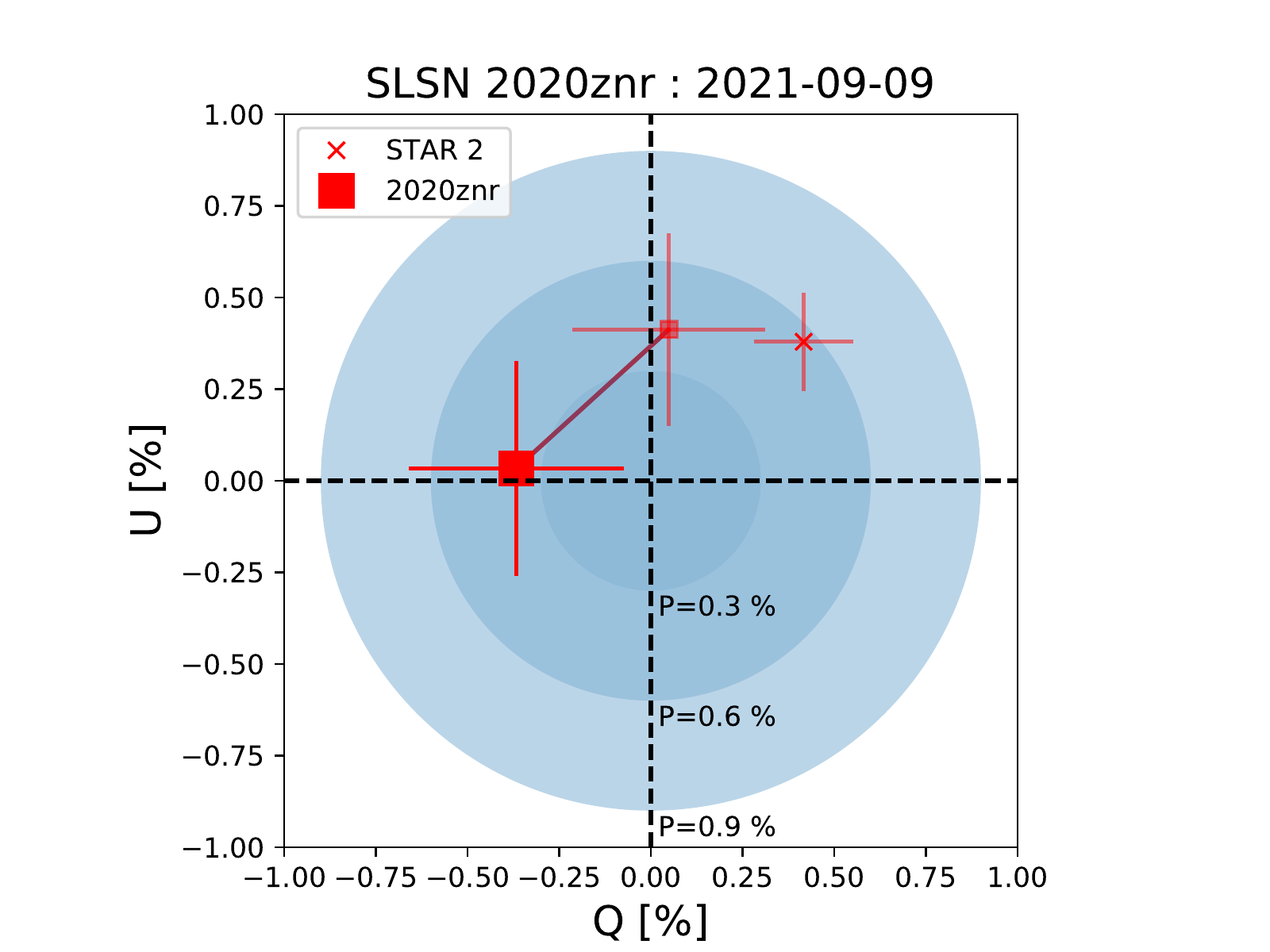}
\vspace*{0.7cm}
\caption{Left column: SN 2020znr, and field stars STAR 1, STAR 2 and STAR 3 
in the $Q-U$ plane, in the plane-of-sky reference frame after 
instrumental polarization (IP) and zero polarization angle (ZPA) 
corrections (see half translucent symbols). 
SN 2020znr Interstellar Polarization (ISP) corrected $Q-U$ estimates have been obtained after subtraction of the IP, ZPA corrected STAR 1, STAR 2 and STAR 3, $Q-U$ weighted average estimates. The values are shown with the double sized square symbols. The double sized square symbols are connected to the half translucent square symbols to show the effect of the ISP correction in the $Q-U$ plane. Right column: same as left column when field star, STAR 2, only is used for estimating the Milky Way ISP contribution. The blue points correspond to the V filter and the red points to the R filter.
  Concentric discs show polarization degrees
ranges up to 0.3 $\%$, 0.6 $\%$ and 0.9 $\%$.}
\label{fig:2020znr_QU_plot}
\end{center}
\end{figure*}

More than 6 months after the first polarimetry measurement, when
SN 2020znr was 1.5 magnitudes fainter, we expected to detect a larger
polarization degree than the one measured during the first polarimetry
epoch. In order to reach a polarization signal-to-noise ratio at least $> 3 \sigma$,
we used a total integration time 4 $\times$ higher in the
R-band only (see log for the first night of September 2021, in Table~\ref{tab:log}), under
two times better seeing conditions. These two factors combined
together lead to a total weighted average $ISP=0.05$ $\pm$ $0.06\%$, i.e. of the same order as on the field stars 
displayed in Table~\ref{tab:heiles}, all about or more
than 8 magnitudes brighter than STAR 1, STAR 2 and STAR 3 
and at heliocentric distances lower than 1000 pc. A level of precision also reached with STAR 2 whose ISP level is $ISP=0.04$ $\pm$ $0.11\%$.
These two ISP estimates lead to a
stringent constraint on the intrinsic level of polarization obtained on SN 2020znr of about 0.1 $\pm$ 0.1$\%$ using STAR 1, STAR 2 and STAR 3, and of about 0.22 $\pm$ 0.14$\%$, using STAR 2 only, i.e. fully consistent with 0$\%$. 
In order to double check this result, the observations repeated on
the following night under similar seeing conditions but with two times
shorter integration times lead to a level of polarization of about
0.4 $\pm$ 0.2$\%$, still consistent with 0$\%$. 
A visual summary of all these results can be seen on 
the $Q - U$ plots displayed in Figure~\ref{fig:2020znr_QU_plot} 
for each polarimetry epoch.

\begin{table*}
	\centering
	\caption{Starlight polarization from the \citet{heiles2000}
          agglomeration file catalog in the vicinity of SN 2020znr.}
	\label{tab:heiles}
	\begin{tabular}{cccccccccccc} 
          \hline
          Star name &  RA (J2000) &  Dec (J2000)  &GLON     &GLAT &  $P$&  $\sigma_{P}$ &  $\theta $ &  $\sigma_{\theta}$ &  V  &  Heliocentric  & Distance to \\
           &   &   &     & &  &   &  &   &   &   Distance &  SN 2020znr \\          
           &   $[^{\circ}]$&  $[^{\circ}]$ & $[^{\circ}]$    & $[^{\circ}]$&  [\%] &   $[\%]$ &  $[^{\circ}]$ &  $[^{\circ}]$ &  [mag] &  [pc] & $[^{\circ}]$ \\
          \hline
          HD 55052 &  108.109950 &      24.1287 &193.2131 & 15.1134&  0.02 &      0.035 &                 38.2 &                          41.2 &      5.7 &            63.1 & 1.67 \\
          HD 57727 &  110.868975 &      25.0506 &193.3412 & 17.7833&  0.04 &      0.120 &                 45.0 &                          56.3 &      5.1 &            50.1 & 1.16 \\
          HD 57702 &  110.852505 &      25.5162 &192.8823 & 17.9453&  0.04 &      0.069 &                  7.3 &                          40.8 &      9.0 &           758.6 & 1.22 \\         
          \hline
	\end{tabular}
\end{table*}

We point out that the object observed in the Legacy survey of magnitude g = 23.96 (see introduction section), very likely the host of SN 2020znr, is very faint with respect to the brightness of SN 2020znr. An intrinsic polarization degree associated to this feature, very likely a blue dwarf galaxy of absolute magnitudes, -14.32 mag in g-band, and -14.64 mag in r-band, assuming a redshift $z=0.1,$ is therefore not expected to contribute to the total polarization degree measured on SN 2020znr. On the other hand, the light emitted by SN 2020znr could be polarized by aligned dust grains pervading such a dwarf galaxy, even though it is a low metallicity galaxy. Dust polarization properties of such galaxies are not known and one can not discard the existence of magnetically aligned dust grains in such environments. For comparison, the low metallicity Magellanic clouds of absolute magnitude of about -18 mag in V-band, are pervaded by complex magnetic fields aligning dust grains at the origin of starlight polarization \citep[e.g.][]{lobo-gomes2015}. Our time-dependent observations can not directly help distinguish an intrinsically null polarization signal from an intrinsic polarization signal from SN 2020znr that is partly cancelled by its host ISP. On one hand, however, if the host ISP is not null it is expected to be uniform with time. On the other hand, comparisons with previous studies (see section~\ref{intro} and section~\ref{discussion}) strongly suggests that the intrinsic polarization of SLSNe are almost null during early phases (see the Introduction section). If SN 2020znr falls into such a category, therefore, one could assume that the ISP from its host is negligible in our detections. This is the hypothesis that will sustain our discussion in the following section.

\section{Discussion} \label{discussion}

Contrary to what has been observed on SN 2015bn a few days after maximum light, and on SN 2017egm about +185 days after maximum light, polarimetry on SN 2020znr does not show an increase of the degree of polarization, departing from null-polarization, at late time epochs ($\approx$ +238 and +239 days). In the following we discuss further comparisons between SN 2020znr, SN 2015bn, and SN 2017egm, that may help to better understand this difference. 

\begin{figure}
\begin{center}
\vspace*{2mm}
\centering
\includegraphics[width=90mm,angle=0]{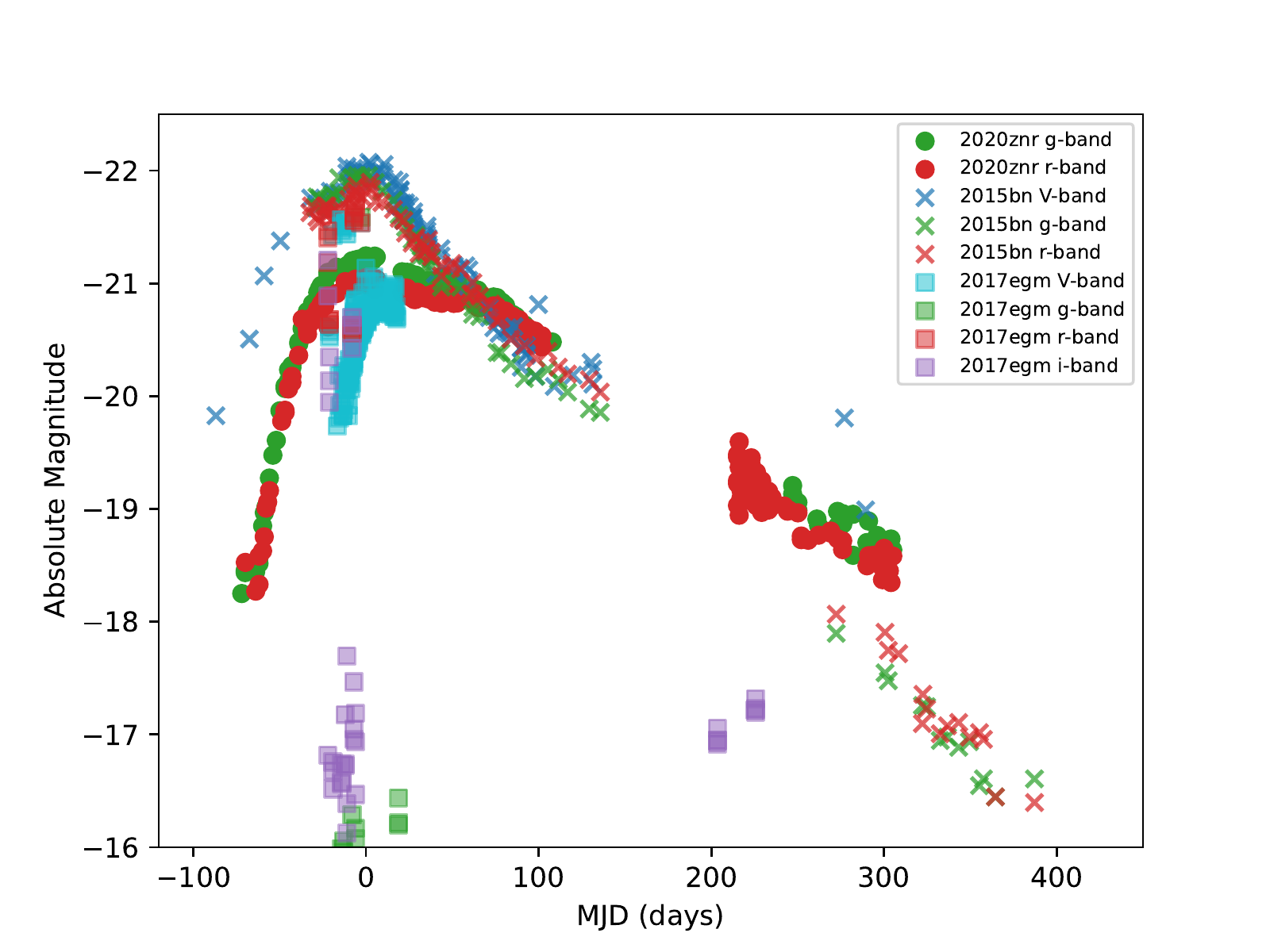}
\vspace*{0.7cm}
\caption{Light curves of SN 2015bn \citep[][]{2016ApJ...826...39N}, SN 2017egm \citep[][]{2017ApJ...845L...8N}, and SN 2020znr (this work, see section~\ref{data_phot}). Data from SN 2015bn and SN 2017egm were retrieved from the Open Supernova Catalog.}
\label{fig:plots_3_SLSNe}
\end{center}
\end{figure}

Figure~\ref{fig:plots_3_SLSNe} shows the absolute magnitude light curves of SN 2020znr, SN 2015bn, and SN 2017egm. The distance modulus was calculated using the \textit{Planck} 2018 Flat $\Lambda-$CDM cosmology model ($\Omega_{0}=0.31, H_{0}=67.7$ km$/$s) \citep[][]{planck2018i}. For SN 2020znr we use a redhsift $z=0.1$ while for SN 2015bn and SN 2017egm we use redshifts $z=0.1136$ and $z=0.030721$, respectively. Photometry on SN 2015bn and SN 2017egm was retrieved from the OSC \footnote{Open Supernova Catalog, {\tt
    https://sne.space}} \citep{2017ApJ...835...64G}. Light curves comparisons show smaller fading timescales for SN 2017egm and SN 2015bn than for SN 2020znr, as well as a shorter rising timescale for SN 2017egm while SN 2020znr and SN 2015bn show similar rising timescales. At maximum light SN 2015bn appears to be more luminous< ($\approx -22$ mag) than SN 2017egm ($\approx -21.6$ magnitudes) and SN 2020znr ($\approx -21.2$ magnitudes).   
Magnetar models of SN 2015bn and SN 2017egm light curves have been explored and discussed by \citet{2017ApJ...850...55N} and \citet{2017ApJ...845L...8N}, respectively. Some of the parameters obtained from these analyses are displayed in Table \ref{tab:mosfitparams} for comparison with the best fit value parameters obtained on SN 2020znr. They show that SN 2017egm, the fastest transient of the three sources, has an estimated ejected mass, $M_{\rm ej}$, of order 2-4 $M_\odot$, while for SN 2015bn and SN 2020znr $M_{\rm ej}$ is of $\approx$ 12 $M_\odot$ and $\approx$ 21 $M_\odot$, respectively. SN 2015bn and SN 2020znr have ejecta velocity of $\approx 5500 $ km$/$s each, and total kinetic energy of the same order, with $E_{\rm K} \approx 4 \times 10^{51}$ ergs for SN 2020znr, and $E_{\rm min, K} \approx 3.45 \times 10^{51}$ ergs for SN 2015bn \citep{2017ApJ...850...55N}, while SN 2017egm has a total kinetic energy of $\approx 1 - 2 \times 10^{51}$ ergs and an ejecta velocity of $\approx 5500 $ km$/$s. All put together, these differences could explain why SN 2017egm is such a quick transient with respect to SN 2015bn and SN 2020znr. In addition, with similar total kinetic energy and ejecta velocity conditions, SN 2020znr has a higher mass ejecta than SN 2015bn. This could make probing any axisymmetry of the photosphere of the inner ejecta of SN 2020znr more difficult, on similar timescales. Which, in turn, could explain why null-polarization is still measured about 238 days after maximum light on SN 2020znr. We note on the other hand, that SN 2020znr do not seem to show any prominent post peak bumps, contrary to what has been quantified and discussed for SN 2015bn and SN 2017egm by \citet{2021arXiv210909743H}. This may also give hints why SN 2020znr does not show any sign of asymmetry at late phases. 

Spectroscopy may be also informative of some differences between the 3 sources. A multi-phase spectral database of SLSNe has been tentatively classified into two classes based on spectra features analysis and template fitting procedures by \citet{2018ApJ...855....2Q}, meaning SN 2020znr may be closer to the PTF12dam type than to the SN 2011ke type. Unfortunately, this analysis does not include spectra from neither SN 2015bn nor SN 2017egm. Another analysis on another sample of spectra of 28 type I SLSNe, including light curves retrieved from the OSC, has been investigated by \citet{2021ApJ...909...24K}. In their work the authors infer ejecta masses via the formalism of diffusion equations, while photospheric velocities were estimated using a method combining spectrum modelling with cross-correlation. The authors find that the W-shaped O II absorption blend, typically present in early phase spectra, is missing from the spectra of several objects having similar features to SN 2015bn. Two classes of object are therefore considered called Type W and Type 15bn. From the spectral analysis, the calculations confirm that Fast rising SLSNe generally show higher photospheric velocities close to maximum than Slow rising events. In this framework, Type 15bn events are considered as Slow evolving events, while Type W events are represented in both the Fast and Slow rising groups. Making a full analysis of the spectra of SN 2020znr is out of the scope of this work, but the comparisons of the spectra of SN 2020znr with those of PFT12dam displayed in Figure~\ref{fig:2020znr_SPEC_AND_FILT} infer SN 2020znr to be part of the Type W group proposed by \citet{2021ApJ...909...24K}. The light curve comparisons displayed in Figure~\ref{fig:plots_3_SLSNe} also infer SN 2020znr to be part of the Slow group, as does SN 2015bn. In other words, SN 2015bn and SN 2020znr could be part of two different spectral classes. On the other hand early spectra obtained on SN 2017egm clearly show that the W signature and should be a fast-rising member of the Type W group. From the sample of SLSNe discussed by \citet{2021ApJ...909...24K}, SN 2018hti and SN LSQ14mo are type I SLSNe members of the Type W group. Polarimetry on SN 2018hti was obtained by \citet{lee2019} before and after maximum light, in a total of 3 different epochs. The results are dominated by the ISP and do not show $> 3 \sigma$ signal-to-noise polarization detection variations. Its light curve is similar to the one of 2015bn \citep{2020MNRAS.497..318L} suggesting it is part of the Slow rising event group. Polarimetry obtained by \citep{leloudas2015} on SN LSQ14mo at 5 different epochs is also found to be consistent with null-polarization detection. Unfortunately, the rising part of the light curve of SN LSQ14mo is not known, making it difficult to classify as part of the Slow or Fast rising group. All in all, these different categorization of SLSNe into several groups (W versus bn15 Type, Slow versus Fast rising event, polarized versus non-polarized system) may be useful to better understand the nature of type I SLSNe, but polarimetry data obtained at early and late epochs would be needed on a larger sample of sources for one to be able to drive any conclusions on a statistical basis. A summary of the comparisons above is given in Table~\ref{tab:comparisons}.  

\begin{table}
	\centering
	\caption{Possible categorization of some SLSNe that have been probed with linear polarimetry. 
	$^{\rm(a)}$: \citet{2021ApJ...909...24K}. 
	$^{\rm(b)}$: This work. 
	$^{\rm(c)}$: \citet{2017ApJ...845L...8N}. 
	$^{\rm(d)}$: \citet{2017ApJ...850...55N}.
    $^{\rm(e)}$: \citep{2020MNRAS.497..318L}
	$^{\rm(f)}$: \citet{inserra2016, leloudas2017}. 
	$^{\rm(g)}$: \citet{saito2020}. 
	$^{\rm(h)}$: \citet{lee2019}.
	$^{\rm(i)}$: \citet{leloudas2015}}
	\label{tab:comparisons}
	\begin{tabular}{llll} 
          \hline
          SLSN &  W / 15bn$^{\rm(a)}$ &  Fast / Slow$^{\rm(a)}$ &  $P$ $^{\rm(b)}$  \\
               &  Type &   rising event &  [\%]   \\
          \hline
          2020znr & W$^{\rm(b)}$    & Slow$^{\rm(b)}$ & null $^{\rm(b)}$\\
          2015bn  & bn15$^{\rm(a)}$ & Slow$^{\rm(d)}$ & increase with time$^{\rm(f)}$ \\
          2017egm & W$^{\rm(c)}$    & Fast$^{\rm(c)}$ & increase with time$^{\rm(g)}$ \\
          2018hti & W$^{\rm(a)}$    & Slow$^{\rm(e)}$ & null$^{\rm(h)}$ \\
          LSQ14mo & W$^{\rm(a)}$    & ?               & null$^{\rm(i)}$ \\
          \hline
	\end{tabular}
\end{table}

Additionally, constraints on jets of SLSN-I from follow-up radio observation analysis as the ones provided by \citet{2018ApJ...856...56C} may be useful to put constraints on the geometry of the inner ejecta, and help to the interpretation of polarization measurements. It is also worth noting that imaging polarimetry and spectropolarimetry may give different polarization diagnostics if the overall electron distribution is mostly spherical, but the abundance of some element may be non-spherical. This might happen here for SN 2020znr which has been observed with broad band linear polarimetry only. SN 2015bn and SN 2017egm were both observed with linear spectropolarimetry (\cite{inserra2016} and \cite{saito2020} respectively), which by definition is less subject to mitigation effect of the net polarization signal due to the integration of the signal over a broader band. The increase of polarization detected on SN 2017egm was from the analysis of linear spectropolarimetry data only \citep{saito2020}. On the other hand the increase of linear polarization observed on SN 2015bn was detected both with linear  spectropolarimetry and broad band linear polarimetry (\citep{inserra2016}.

\section{Summary}

Optical imaging polarimetry was conducted on the hydrogen poor SLSN 2020znr during 3 phases after maximum light ($\approx$ +34 days, +288 days and +289 days with respect to g-band). After instrumental and interstellar polarization correction, all measurements are consistent with null-polarization detection.

The light curve including ZTF g- and r-band, and ATLAS c- and o-band data has been modelled with \texttt{MOSFiT}. The best fit values from the SLSN model displayed in Table \ref{tab:mosfitparams} show that most of the characteristics of SN 2020znr are not significantly different from other SLSNe in this parameter space.

A comparison of the \texttt{MOSFiT} best-fit values discussed in the literature on SN 2017egm and SN 2015bn, two hydrogen poor SLSNe showing an increase of polarization after maximum light, suggests that SN 2020znr has an higher mass ejecta that may prevent access to the geometry of the inner ejecta with optical polarimetry.

The combined information provided by spectroscopy and light curve analysis may be helpful to distinguish several classes of type I SLSNe showing different polarization signatures. Such an avenue could be explored with a larger sample of SLSNe polarimetry data.  


\section*{Acknowledgements}

The authors would like to thank the anonymous referee for helpful comments, Matt Nicholl and Mattia Bulla for their helpful discussion, Hannu Parviainen for his kind support on the NOT, Sergio Armas Perez for providing data from the NOT public archive, and David Young for the ATLAS python public code he developed to stack and bin ATLAS forced photometry public data.  

F.P. acknowledges support from the Spanish State Research Agency (AEI) under grant number PID2019-105552RB-C43. I.P.-F. acknowledges support from the Spanish State Research Agency (AEI) under grant numbers ESP2017-86852-C4-2-R and PID2019-105552RB- C43. 

Based on observations made with the Nordic Optical Telescope (NOT), owned in collaboration by the University of Turku and Aarhus University, and operated jointly by Aarhus University, the University of Turku and the University of Oslo, representing Denmark, Finland and Norway, the University of Iceland and Stockholm University at the Observatorio del Roque de los Muchachos, La Palma, Spain, of the Instituto de Astrofisica de Canarias. The data presented here were obtained in part with ALFOSC, which is provided by the Instituto de Astrofisica de Andalucia (IAA) under a joint agreement with the University of Copenhagen and NOT. Some of the data were obtained during CAT service observation Spanish time. ALFOSC polarimetry imaging data of SN 2020ank and calibration data were retrieved from the NOT public archive. 

The Liverpool Telescope is operated on the island of La Palma by Liverpool John Moores University in the Spanish Observatorio del Roque de los Muchachos of the Instituto de Astrofisica de Canarias with financial support from the UK Science and Technology Facilities Council.

This work is based in part on observations obtained with the Samuel Oschin 48-inch Telescope at the Palomar Observatory as part of the Zwicky Transient Facility project. ZTF is supported by the NSF under grant AST-1440341 and a collaboration including Caltech, IPAC, the Weizmann Institute for Science, the Oskar Klein Center at Stockholm University, the University of Maryland, the University of Washington, Deutsches Elektronen-Synchrotron and Humboldt University, Los Alamos National Laboratories, the TANGO Consortium of Taiwan, the University of Wisconsin at Milwaukee, and the Lawrence Berkeley National Laboratory. Operations are conducted by the Caltech Optical Observatories (COO), the Infrared Processing and Analysis Center (IPAC), and the University of Washington (UW).

This work has made use of data from the Asteroid Terrestrial- impact Last Alert System (ATLAS) project. The Asteroid Terrestrial- impact Last Alert System (ATLAS) project is primarily funded to search for near earth asteroids through NASA grants NN12AR55G, 80NSSC18K0284, and 80NSSC18K1575; byproducts of the NEO search include images and catalogs from the survey area. This work was partially funded by Kepler/K2 grant J1944/80NSSC19K0112 and HST GO-15889, and STFC grants ST/T000198/1 and ST/S006109/1. The ATLAS science products have been made possible through the contributions of the University of Hawaii Institute for Astronomy, the Queen s University Belfast, the Space Telescope Science Institute, the South African Astronomical Observatory, and The Millennium Institute of Astrophysics (MAS), Chile.

Lasair is supported by the UKRI Science and Technology Facilities Council and is a collaboration between the University of Edinburgh (grant ST/N002512/1) and Queen’s University Belfast (grant ST/N002520/1) within the LSST:UK Science Consortium.

This research has made use of ``Aladin sky atlas'' developed at CDS, Strasbourg Observatory, France 2000A\&AS..143...33B and 2014ASPC..485..277B.

SNID is Copyright (C) 1999-2007 St\'{e}phane Blondin and John L. Tonry, and is available under the GNU General Public License.

This work made use of the python public code \texttt{astropy/photutils: 1.0.2} release developed by \citet{2021zndo...4453725B}.

This research made use of the Transient Name Server (TNS) which is the official IAU mechanism for reporting new astronomical transients such as supernova candidates, As of January 1, 2016.

This research made use of DESI LS DR9 data. The Legacy Surveys consist of three individual and complementary projects: the Dark Energy Camera Legacy Survey (DECaLS; Proposal ID \#2014B-0404; PIs: David Schlegel and Arjun Dey), the Beijing-Arizona Sky Survey (BASS; NOAO Prop. ID \#2015A-0801; PIs: Zhou Xu and Xiaohui Fan), and the Mayall z-band Legacy Survey (MzLS; Prop. ID \#2016A-0453; PI: Arjun Dey). DECaLS, BASS and MzLS together include data obtained, respectively, at the Blanco telescope, Cerro Tololo Inter-American Observatory, NSF’s NOIRLab; the Bok telescope, Steward Observatory, University of Arizona; and the Mayall telescope, Kitt Peak National Observatory, NOIRLab. The Legacy Surveys project is honored to be permitted to conduct astronomical research on Iolkam Du’ag (Kitt Peak), a mountain with particular significance to the Tohono O’odham Nation.

NOIRLab is operated by the Association of Universities for Research in Astronomy (AURA) under a cooperative agreement with the National Science Foundation.

This project used data obtained with the Dark Energy Camera (DECam), which was constructed by the Dark Energy Survey (DES) collaboration. Funding for the DES Projects has been provided by the U.S. Department of Energy, the U.S. National Science Foundation, the Ministry of Science and Education of Spain, the Science and Technology Facilities Council of the United Kingdom, the Higher Education Funding Council for England, the National Center for Supercomputing Applications at the University of Illinois at Urbana-Champaign, the Kavli Institute of Cosmological Physics at the University of Chicago, Center for Cosmology and Astro-Particle Physics at the Ohio State University, the Mitchell Institute for Fundamental Physics and Astronomy at Texas A$\&$M University, Financiadora de Estudos e Projetos, Fundacao Carlos Chagas Filho de Amparo, Financiadora de Estudos e Projetos, Fundacao Carlos Chagas Filho de Amparo a Pesquisa do Estado do Rio de Janeiro, Conselho Nacional de Desenvolvimento Cientifico e Tecnologico and the Ministerio da Ciencia, Tecnologia e Inovacao, the Deutsche Forschungsgemeinschaft and the Collaborating Institutions in the Dark Energy Survey. The Collaborating Institutions are Argonne National Laboratory, the University of California at Santa Cruz, the University of Cambridge, Centro de Investigaciones Energeticas, Medioambientales y Tecnologicas-Madrid, the University of Chicago, University College London, the DES-Brazil Consortium, the University of Edinburgh, the Eidgenossische Technische Hochschule (ETH) Zurich, Fermi National Accelerator Laboratory, the University of Illinois at Urbana-Champaign, the Institut de Ciencies de l’Espai (IEEC/CSIC), the Institut de Fisica d’Altes Energies, Lawrence Berkeley National Laboratory, the Ludwig Maximilians Universitat Munchen and the associated Excellence Cluster Universe, the University of Michigan, NSF’s NOIRLab, the University of Nottingham, the Ohio State University, the University of Pennsylvania, the University of Portsmouth, SLAC National Accelerator Laboratory, Stanford University, the University of Sussex, and Texas A$\&$M University.

The Legacy Surveys imaging of the DESI footprint is supported by the Director, Office of Science, Office of High Energy Physics of the U.S. Department of Energy under Contract No. DE-AC02-05CH1123, by the National Energy Research Scientific Computing Center, a DOE Office of Science User Facility under the same contract; and by the U.S. National Science Foundation, Division of Astronomical Sciences under Contract No. AST-0950945 to NOAO.

\section*{Data Availability}

For science reproducibility purposes, the photometry data displayed in section~\ref{photometry_table} will be available online.
The ESO-NTT $/$ EFOSC2 and the LT $/$ SPRAT spectra presented in this work are available via WISEReP.  




\bibliographystyle{mnras}
\bibliography{sn2020znr} 




\appendix

\section{Test on our Polarization Data Reduction Pipeline} \label{pipeline}

We ran our polarimetry data reduction pipeline on the ALFOSC archival data obtained on SN 2020ank by \citet{lee2020}.
Our results are displayed in Table~\ref{tab:pipeline} for comparison.
Note that polarized star HD251204 reported in Table 2 from \citet{lee2020} is actually polarized star HD236928, as from the coordinates displayed in the fits header of the ALFOSC dataframes. The outputs obtained with the two methods are consistent with each other within the uncertainties.

\begin{table}
	\centering
	\caption{Polarimetry results obtained in this work and by \citet{lee2020} after data reduction of the ALFOSC data obtained on March 1, 2020 on SN 2020ank. $^{\rm(a)}$: Highly polarized standard star. $^{\rm(b)}$: unpolarized standard star.}
	\label{tab:pipeline}
	\begin{tabular}{lcc} 
          \hline
          Object & $P$[\%] &   $P$[\%]   \\
                 & \citet{lee2020} &   This work   \\
          \hline
          2020ank           & 0.6 $\pm$ 0.3  & 0.5 $\pm$ 0.2 \\
          HD236928$^{\rm(a)}$ & 6.8 $\pm$ 0.1  & 6.7 $\pm$ 0.1 \\
          HD64299$^{\rm(b)}$  & 0.1 $\pm$ 0.1  & 0.1 $\pm$ 0.1 \\
          \hline
	\end{tabular}
\end{table}


\section{Photometry Table} \label{photometry_table}

The photometry from SN 2020znr used along this work is compiled in
Table~\ref{tab:photometry1}, Table~\ref{tab:photometry2},
Table~\ref{tab:photometry3}, Table~\ref{tab:photometry4},
Table~\ref{tab:photometry5} and Table~\ref{tab:photometry6}.

\begin{table}
	\centering
	\caption{ZTF and ATLAS photometry.}
	\label{tab:photometry1}
	\begin{tabular}{lcccccc} 
          \hline
          MJD &
                Magnitude &  $\sigma_{\rm Magnitude}$& band & survey \\
          \hline
 59145.600000 &    22.620 &       3.554 &      o &     ATLAS \\
 59147.580000 &    21.677 &       1.532 &      c &     ATLAS \\
 59153.560000 &    21.827 &       3.241 &      o &     ATLAS \\
 59155.411944 &    20.034 &       0.314 &      g &       ZTF \\
 59157.399086 &    19.848 &       0.328 &      g &       ZTF \\
 59157.399549 &    19.832 &       0.352 &      g &       ZTF \\
 59157.479387 &    19.758 &       0.282 &      r &       ZTF \\
 59163.440486 &    19.840 &       0.150 &      g &       ZTF \\
 59163.503276 &    20.012 &       0.152 &      r &       ZTF \\
 59163.560000 &    20.312 &       0.482 &      o &     ATLAS \\
 59165.393819 &    19.768 &       0.118 &      g &       ZTF \\
 59165.421620 &    19.702 &       0.127 &      r &       ZTF \\
 59165.422083 &    19.953 &       0.151 &      r &       ZTF \\
 59165.444097 &    19.710 &       0.113 &      g &       ZTF \\
 59167.470729 &    19.656 &       0.132 &      r &       ZTF \\
 59167.506516 &    19.434 &       0.116 &      g &       ZTF \\
 59168.436412 &    19.531 &       0.112 &      r &       ZTF \\
 59168.478322 &    19.317 &       0.087 &      g &       ZTF \\
 59169.468345 &    19.265 &       0.089 &      g &       ZTF \\
 59169.505590 &    19.277 &       0.123 &      r &       ZTF \\
 59170.404248 &    19.222 &       0.086 &      r &       ZTF \\
 59171.416829 &    19.009 &       0.068 &      g &       ZTF \\
 59171.477951 &    19.121 &       0.096 &      r &       ZTF \\
 59173.443137 &    18.807 &       0.096 &      g &       ZTF \\
 59175.421458 &    18.676 &       0.199 &      g &       ZTF \\
 59175.620000 &    18.682 &       0.112 &      c &     ATLAS \\
 59177.516759 &    18.413 &       0.063 &      g &       ZTF \\
 59178.540741 &    18.506 &       0.066 &      r &       ZTF \\
 59179.520000 &    18.558 &       0.114 &      o &     ATLAS \\
 59180.451736 &    18.409 &       0.079 &      r &       ZTF \\
 59180.496250 &    18.215 &       0.079 &      g &       ZTF \\
 59180.496724 &    18.193 &       0.067 &      g &       ZTF \\
 59180.539271 &    18.427 &       0.138 &      r &       ZTF \\
 59181.340000 &    18.419 &       0.319 &      o &     ATLAS \\
 59182.463900 &    18.200 &       0.072 &      r &       ZTF \\
 59182.464363 &    18.220 &       0.058 &      r &       ZTF \\
 59182.529329 &    18.147 &       0.082 &      g &       ZTF \\
 59182.529792 &    18.050 &       0.080 &      g &       ZTF \\
 59183.420000 &    18.070 &       0.172 &      o &     ATLAS \\
 59184.429444 &    18.032 &       0.063 &      g &       ZTF \\
 59184.436053 &    18.012 &       0.071 &      g &       ZTF \\
 59184.464363 &    18.163 &       0.069 &      r &       ZTF \\
 59184.480845 &    18.108 &       0.077 &      r &       ZTF \\
 59188.368275 &    17.922 &       0.059 &      r &       ZTF \\
 59188.445347 &    17.803 &       0.053 &      g &       ZTF \\
 59188.445810 &    17.819 &       0.053 &      g &       ZTF \\
 59189.470000 &    17.997 &       0.094 &      o &     ATLAS \\
 59190.370405 &    17.685 &       0.047 &      g &       ZTF \\
 59190.436458 &    17.601 &       0.231 &      r &       ZTF \\
 59191.460000 &    17.824 &       0.058 &      o &     ATLAS \\
 59192.480000 &    17.627 &       0.043 &      c &     ATLAS \\
 59193.308009 &    17.684 &       0.052 &      r &       ZTF \\
 59193.357211 &    17.545 &       0.037 &      g &       ZTF \\
 59193.434907 &    17.532 &       0.033 &      g &       ZTF \\
 59193.499178 &    17.733 &       0.061 &      r &       ZTF \\
 59193.550000 &    17.756 &       0.044 &      o &     ATLAS \\
 59195.479213 &    17.629 &       0.048 &      r &       ZTF \\
 59195.560000 &    17.522 &       0.036 &      c &     ATLAS \\
 59196.394317 &    17.582 &       0.038 &      r &       ZTF \\
 59196.423449 &    17.462 &       0.053 &      g &       ZTF \\
 59197.410000 &    17.461 &       0.035 &      c &     ATLAS \\
 59198.429722 &    17.593 &       0.043 &      r &       ZTF \\
           \hline
	\end{tabular}
\end{table}

\begin{table}
	\centering
	\caption{ZTF and ATLAS photometry.}
	\label{tab:photometry2}
	\begin{tabular}{lcccccc} 
          \hline
          MJD &
                Magnitude &  $\sigma_{\rm Magnitude}$& band & survey \\
          \hline
 59199.299664 &    17.516 &       0.050 &      r &       ZTF \\
 59199.424861 &    17.364 &       0.028 &      g &       ZTF \\
 59200.342558 &    17.536 &       0.081 &      r &       ZTF \\
 59200.432095 &    17.331 &       0.050 &      g &       ZTF \\
 59201.412697 &    17.302 &       0.029 &      g &       ZTF \\
 59201.458252 &    17.464 &       0.037 &      r &       ZTF \\
 59203.313333 &    17.304 &       0.031 &      g &       ZTF \\
 59203.344803 &    17.294 &       0.038 &      g &       ZTF \\
 59203.431690 &    17.486 &       0.051 &      r &       ZTF \\
 59203.433588 &    17.421 &       0.037 &      r &       ZTF \\
 59205.215961 &    17.394 &       0.041 &      r &       ZTF \\
 59205.301782 &    17.186 &       0.032 &      g &       ZTF \\
 59205.319757 &    17.670 &       0.050 &      r &       ZTF \\
 59205.380776 &    17.209 &       0.050 &      g &       ZTF \\
 59205.520000 &    17.299 &       0.029 &      c &     ATLAS \\
 59207.430000 &    17.400 &       0.052 &      o &     ATLAS \\
 59209.420000 &    17.380 &       0.056 &      o &     ATLAS \\
 59210.263762 &    17.140 &       0.043 &      g &       ZTF \\
 59210.387604 &    17.374 &       0.043 &      r &       ZTF \\
 59215.297755 &    17.268 &       0.075 &      r &       ZTF \\
 59215.337269 &    17.151 &       0.064 &      g &       ZTF \\
 59216.316215 &    17.318 &       0.041 &      r &       ZTF \\
 59216.423947 &    17.125 &       0.059 &      g &       ZTF \\
 59217.325949 &    17.287 &       0.037 &      r &       ZTF \\
 59217.520000 &    17.398 &       0.050 &      o &     ATLAS \\
 59218.265845 &    17.116 &       0.042 &      g &       ZTF \\
 59218.402974 &    17.330 &       0.042 &      r &       ZTF \\
 59219.279317 &    17.084 &       0.027 &      g &       ZTF \\
 59219.302188 &    17.269 &       0.037 &      r &       ZTF \\
 59219.490000 &    17.350 &       0.047 &      o &     ATLAS \\
 59220.297824 &    17.303 &       0.040 &      r &       ZTF \\
 59220.339352 &    17.093 &       0.044 &      g &       ZTF \\
 59221.234606 &    17.246 &       0.040 &      r &       ZTF \\
 59221.318472 &    17.075 &       0.033 &      g &       ZTF \\
 59221.410000 &    17.171 &       0.027 &      c &     ATLAS \\
 59222.338796 &    17.302 &       0.052 &      r &       ZTF \\
 59222.382951 &    17.095 &       0.034 &      g &       ZTF \\
 59223.296285 &    17.257 &       0.039 &      r &       ZTF \\
 59223.351829 &    17.067 &       0.034 &      g &       ZTF \\
 59223.480000 &    17.202 &       0.070 &      c &     ATLAS \\
 59224.288681 &    17.278 &       0.036 &      r &       ZTF \\
 59225.238461 &    17.078 &       0.039 &      g &       ZTF \\
 59226.387257 &    17.075 &       0.040 &      g &       ZTF \\
 59227.297222 &    17.233 &       0.040 &      r &       ZTF \\
 59227.342107 &    17.040 &       0.038 &      g &       ZTF \\
 59227.600000 &    17.148 &       0.031 &      c &     ATLAS \\
 59228.265185 &    17.090 &       0.037 &      g &       ZTF \\
 59228.318634 &    17.302 &       0.039 &      r &       ZTF \\
 59229.221076 &    17.100 &       0.041 &      g &       ZTF \\
 59229.333032 &    17.259 &       0.038 &      r &       ZTF \\
 59229.470000 &    17.144 &       0.031 &      c &     ATLAS \\
 59230.244039 &    17.257 &       0.042 &      r &       ZTF \\
 59230.328750 &    17.075 &       0.051 &      g &       ZTF \\
 59231.234178 &    17.081 &       0.031 &      g &       ZTF \\
 59231.311794 &    17.251 &       0.038 &      r &       ZTF \\
 59231.360000 &    17.127 &       0.029 &      c &     ATLAS \\
 59232.372373 &    17.339 &       0.058 &      r &       ZTF \\
 59232.394526 &    17.042 &       0.040 &      g &       ZTF \\
 59233.277859 &    17.243 &       0.044 &      r &       ZTF \\
 59233.323391 &    17.049 &       0.027 &      g &       ZTF \\
 59235.430000 &    17.350 &       0.071 &      o &     ATLAS \\
 59245.430000 &    17.398 &       0.051 &      o &     ATLAS \\
           \hline
	\end{tabular}
\end{table}

\begin{table}
	\centering
	\caption{ZTF and ATLAS photometry.}
	\label{tab:photometry3}
	\begin{tabular}{lcccccc} 
          \hline
          MJD &
                Magnitude &  $\sigma_{\rm Magnitude}$& band & survey \\
          \hline
 59248.228542 &    17.181 &       0.042 &      g &       ZTF \\
 59248.236898 &    17.367 &       0.037 &      r &       ZTF \\
 59249.219618 &    17.208 &       0.071 &      g &       ZTF \\
 59249.310660 &    17.325 &       0.039 &      r &       ZTF \\
 59250.256655 &    17.222 &       0.040 &      g &       ZTF \\
 59250.328484 &    17.389 &       0.046 &      r &       ZTF \\
 59251.216262 &    17.326 &       0.034 &      r &       ZTF \\
 59251.275498 &    17.185 &       0.028 &      g &       ZTF \\
 59251.410000 &    17.265 &       0.035 &      c &     ATLAS \\
 59252.204317 &    17.397 &       0.045 &      r &       ZTF \\
 59252.287187 &    17.213 &       0.036 &      g &       ZTF \\
 59253.199132 &    17.199 &       0.035 &      g &       ZTF \\
 59253.297789 &    17.357 &       0.044 &      r &       ZTF \\
 59253.360000 &    17.268 &       0.032 &      c &     ATLAS \\
 59254.202639 &    17.415 &       0.050 &      r &       ZTF \\
 59254.227951 &    17.206 &       0.039 &      g &       ZTF \\
 59255.186354 &    17.232 &       0.028 &      g &       ZTF \\
 59255.297766 &    17.427 &       0.074 &      r &       ZTF \\
 59255.390000 &    17.303 &       0.034 &      c &     ATLAS \\
 59256.213113 &    17.426 &       0.052 &      r &       ZTF \\
 59256.296319 &    17.211 &       0.050 &      g &       ZTF \\
 59257.214421 &    17.363 &       0.038 &      r &       ZTF \\
 59257.390000 &    17.252 &       0.033 &      c &     ATLAS \\
 59260.192523 &    17.262 &       0.047 &      g &       ZTF \\
 59260.249977 &    17.410 &       0.039 &      r &       ZTF \\
 59261.340000 &    17.245 &       0.030 &      c &     ATLAS \\
 59262.191921 &    17.396 &       0.043 &      r &       ZTF \\
 59262.276562 &    17.260 &       0.028 &      g &       ZTF \\
 59263.213750 &    17.420 &       0.051 &      r &       ZTF \\
 59263.277813 &    17.260 &       0.049 &      g &       ZTF \\
 59264.191111 &    17.389 &       0.039 &      r &       ZTF \\
 59264.282326 &    17.266 &       0.038 &      g &       ZTF \\
 59265.187778 &    17.298 &       0.042 &      g &       ZTF \\
 59265.212002 &    17.427 &       0.043 &      r &       ZTF \\
 59265.330000 &    17.594 &       0.961 &      o &     ATLAS \\
 59266.133750 &    17.260 &       0.055 &      g &       ZTF \\
 59266.212928 &    17.376 &       0.041 &      r &       ZTF \\
 59267.174965 &    17.450 &       0.042 &      r &       ZTF \\
 59267.253657 &    17.283 &       0.051 &      g &       ZTF \\
 59270.233553 &    17.273 &       0.052 &      g &       ZTF \\
 59270.295313 &    17.398 &       0.066 &      r &       ZTF \\
 59271.225451 &    17.225 &       0.036 &      g &       ZTF \\
 59271.258333 &    17.460 &       0.053 &      r &       ZTF \\
 59272.170428 &    17.279 &       0.048 &      g &       ZTF \\
 59272.214144 &    17.365 &       0.047 &      r &       ZTF \\
 59273.173437 &    17.249 &       0.048 &      g &       ZTF \\
 59273.212153 &    17.436 &       0.050 &      r &       ZTF \\
 59274.166250 &    17.325 &       0.054 &      g &       ZTF \\
 59274.256100 &    17.396 &       0.067 &      r &       ZTF \\
 59275.172106 &    17.277 &       0.041 &      g &       ZTF \\
 59276.211238 &    17.294 &       0.025 &      g &       ZTF \\
 59276.282407 &    17.382 &       0.049 &      r &       ZTF \\
 59277.340000 &    17.322 &       0.029 &      c &     ATLAS \\
 59278.185938 &    17.461 &       0.046 &      r &       ZTF \\
 59280.150150 &    17.395 &       0.045 &      r &       ZTF \\
 59280.150150 &    17.395 &       0.045 &      r &       ZTF \\
 59280.152998 &    17.457 &       0.047 &      r &       ZTF \\
 59280.152998 &    17.457 &       0.047 &      r &       ZTF \\
 59280.210995 &    17.298 &       0.049 &      g &       ZTF \\
 59280.210995 &    17.298 &       0.049 &      g &       ZTF \\
 59280.211470 &    17.273 &       0.027 &      g &       ZTF \\
 59280.211470 &    17.273 &       0.027 &      g &       ZTF \\
           \hline
	\end{tabular}
\end{table}

\begin{table}
	\centering
	\caption{ZTF and ATLAS photometry.}
	\label{tab:photometry4}
	\begin{tabular}{lcccccc} 
          \hline
          MJD &
                Magnitude &  $\sigma_{\rm Magnitude}$& band & survey \\
          \hline
 59287.300000 &    17.349 &       0.042 &      c &     ATLAS \\
 59290.213958 &    17.383 &       0.048 &      g &       ZTF \\
 59290.214421 &    17.336 &       0.036 &      g &       ZTF \\
 59290.239167 &    17.465 &       0.043 &      r &       ZTF \\
 59290.239630 &    17.429 &       0.041 &      r &       ZTF \\
 59291.360000 &    17.385 &       0.040 &      c &     ATLAS \\
 59292.173160 &    17.366 &       0.044 &      g &       ZTF \\
 59292.174120 &    17.344 &       0.044 &      g &       ZTF \\
 59292.196875 &    17.460 &       0.041 &      r &       ZTF \\
 59292.214398 &    17.499 &       0.054 &      r &       ZTF \\
 59293.310000 &    17.518 &       0.041 &      o &     ATLAS \\
 59294.170185 &    17.385 &       0.040 &      g &       ZTF \\
 59294.208137 &    17.496 &       0.047 &      r &       ZTF \\
 59299.350000 &    17.794 &       1.354 &      o &     ATLAS \\
 59301.177975 &    17.573 &       0.048 &      r &       ZTF \\
 59301.178449 &    17.483 &       0.034 &      r &       ZTF \\
 59301.215012 &    17.404 &       0.051 &      g &       ZTF \\
 59301.234606 &    17.442 &       0.048 &      g &       ZTF \\
 59303.156852 &    17.487 &       0.040 &      r &       ZTF \\
 59303.196296 &    17.423 &       0.047 &      g &       ZTF \\
 59303.218426 &    17.410 &       0.046 &      g &       ZTF \\
 59303.242743 &    17.565 &       0.052 &      r &       ZTF \\
 59305.350000 &    17.537 &       0.055 &      o &     ATLAS \\
 59306.157986 &    17.451 &       0.037 &      g &       ZTF \\
 59306.181944 &    17.477 &       0.042 &      g &       ZTF \\
 59306.195255 &    17.545 &       0.043 &      r &       ZTF \\
 59306.226482 &    17.584 &       0.047 &      r &       ZTF \\
 59308.159514 &    17.479 &       0.036 &      g &       ZTF \\
 59308.186435 &    17.529 &       0.057 &      r &       ZTF \\
 59308.186910 &    17.603 &       0.047 &      r &       ZTF \\
 59308.247130 &    17.494 &       0.042 &      g &       ZTF \\
 59310.184317 &    17.617 &       0.046 &      r &       ZTF \\
 59310.184780 &    17.579 &       0.045 &      r &       ZTF \\
 59312.172523 &    17.632 &       0.043 &      r &       ZTF \\
 59312.172986 &    17.575 &       0.038 &      r &       ZTF \\
 59312.206262 &    17.579 &       0.045 &      g &       ZTF \\
 59312.206736 &    17.558 &       0.048 &      g &       ZTF \\
 59313.290000 &    17.588 &       0.039 &      c &     ATLAS \\
 59314.184826 &    17.639 &       0.049 &      r &       ZTF \\
 59314.205949 &    17.577 &       0.048 &      g &       ZTF \\
 59314.206424 &    17.614 &       0.046 &      g &       ZTF \\
 59316.160012 &    17.683 &       0.048 &      r &       ZTF \\
 59316.160475 &    17.607 &       0.053 &      r &       ZTF \\
 59316.207026 &    17.631 &       0.043 &      g &       ZTF \\
 59316.207500 &    17.632 &       0.051 &      g &       ZTF \\
 59319.174988 &    17.652 &       0.053 &      g &       ZTF \\
 59319.175463 &    17.684 &       0.033 &      g &       ZTF \\
 59319.270000 &    17.710 &       0.057 &      o &     ATLAS \\
 59321.163808 &    17.688 &       0.054 &      g &       ZTF \\
 59321.174248 &    17.697 &       0.056 &      g &       ZTF \\
 59321.200521 &    17.763 &       0.052 &      r &       ZTF \\
 59321.220972 &    17.681 &       0.054 &      r &       ZTF \\
 59321.250000 &    17.778 &       0.058 &      o &     ATLAS \\
 59325.185799 &    17.706 &       0.052 &      r &       ZTF \\
 59325.186262 &    17.809 &       0.071 &      r &       ZTF \\
 59325.199213 &    17.730 &       0.049 &      g &       ZTF \\
 59325.199676 &    17.774 &       0.059 &      g &       ZTF \\
 59327.240000 &    17.815 &       0.108 &      o &     ATLAS \\
 59329.166956 &    17.779 &       0.055 &      g &       ZTF \\
 59329.166956 &    17.779 &       0.055 &      g &       ZTF \\
 59329.198681 &    17.845 &       0.070 &      r &       ZTF \\
 59329.198681 &    17.845 &       0.070 &      r &       ZTF \\
           \hline
	\end{tabular}
\end{table}

\begin{table}
	\centering
	\caption{ZTF and ATLAS photometry.}
	\label{tab:photometry5}
	\begin{tabular}{lcccccc} 
          \hline
          MJD &
                Magnitude &  $\sigma_{\rm Magnitude}$& band & survey \\
          \hline
 59329.199630 &    17.749 &       0.062 &      r &       ZTF \\
 59329.199630 &    17.749 &       0.062 &      r &       ZTF \\
 59329.280000 &    17.755 &       0.111 &      o &     ATLAS \\
 59331.290000 &    17.892 &       0.267 &      o &     ATLAS \\
 59335.217083 &    17.803 &       0.073 &      g &       ZTF \\
 59337.270000 &    17.905 &       0.062 &      o &     ATLAS \\
 59345.260000 &    17.912 &       0.085 &      o &     ATLAS \\
 59347.260000 &    17.937 &       0.081 &      c &     ATLAS \\
 59442.499815 &    19.253 &       0.232 &      r &       ZTF \\
 59442.502199 &    19.035 &       0.197 &      r &       ZTF \\
 59442.504595 &    18.824 &       0.188 &      r &       ZTF \\
 59442.506991 &    19.059 &       0.262 &      r &       ZTF \\
 59442.509387 &    18.803 &       0.233 &      r &       ZTF \\
 59443.498819 &    19.339 &       0.281 &      r &       ZTF \\
 59443.501215 &    19.247 &       0.299 &      r &       ZTF \\
 59443.503611 &    18.916 &       0.273 &      r &       ZTF \\
 59443.508403 &    18.688 &       0.239 &      r &       ZTF \\
 59444.499178 &    19.190 &       0.215 &      r &       ZTF \\
 59444.501574 &    19.076 &       0.221 &      r &       ZTF \\
 59444.503970 &    19.116 &       0.235 &      r &       ZTF \\
 59446.502002 &    19.071 &       0.174 &      r &       ZTF \\
 59446.504363 &    19.147 &       0.183 &      r &       ZTF \\
 59446.506725 &    19.151 &       0.214 &      r &       ZTF \\
 59446.509097 &    19.143 &       0.268 &      r &       ZTF \\
 59449.506644 &    18.839 &       0.293 &      r &       ZTF \\
 59449.509051 &    18.888 &       0.282 &      r &       ZTF \\
 59450.493831 &    18.896 &       0.241 &      r &       ZTF \\
 59450.498669 &    18.966 &       0.268 &      r &       ZTF \\
 59450.501099 &    18.830 &       0.217 &      r &       ZTF \\
 59450.505914 &    18.952 &       0.326 &      r &       ZTF \\
 59450.508310 &    18.935 &       0.241 &      r &       ZTF \\
 59450.513113 &    18.973 &       0.316 &      r &       ZTF \\
 59451.490984 &    19.053 &       0.172 &      r &       ZTF \\
 59451.510382 &    19.081 &       0.187 &      r &       ZTF \\
 59451.512778 &    19.191 &       0.177 &      r &       ZTF \\
 59452.494919 &    19.045 &       0.167 &      r &       ZTF \\
 59452.499861 &    19.115 &       0.191 &      r &       ZTF \\
 59452.502338 &    19.108 &       0.118 &      r &       ZTF \\
 59453.491319 &    18.959 &       0.111 &      r &       ZTF \\
 59453.493796 &    19.248 &       0.166 &      r &       ZTF \\
 59453.496262 &    19.144 &       0.124 &      r &       ZTF \\
 59453.498727 &    19.256 &       0.138 &      r &       ZTF \\
 59453.501192 &    19.163 &       0.157 &      r &       ZTF \\
 59454.493171 &    19.099 &       0.141 &      r &       ZTF \\
 59454.495648 &    19.178 &       0.172 &      r &       ZTF \\
 59454.498113 &    19.203 &       0.144 &      r &       ZTF \\
 59454.503067 &    19.155 &       0.148 &      r &       ZTF \\
 59455.620000 &    18.987 &       0.344 &      o &     ATLAS \\
 59456.495729 &    19.142 &       0.154 &      r &       ZTF \\
 59456.498241 &    19.033 &       0.124 &      r &       ZTF \\
 59456.500764 &    19.314 &       0.155 &      r &       ZTF \\
 59456.505810 &    19.057 &       0.130 &      r &       ZTF \\
 59460.495660 &    19.275 &       0.123 &      r &       ZTF \\
 59460.498194 &    19.293 &       0.115 &      r &       ZTF \\
 59460.500729 &    19.127 &       0.105 &      r &       ZTF \\
 59460.503264 &    19.168 &       0.114 &      r &       ZTF \\
 59460.505799 &    19.241 &       0.120 &      r &       ZTF \\
 59461.494711 &    19.263 &       0.112 &      r &       ZTF \\
 59461.497245 &    19.245 &       0.115 &      r &       ZTF \\
 59461.499780 &    19.248 &       0.112 &      r &       ZTF \\
 59461.502315 &    19.260 &       0.126 &      r &       ZTF \\
 59461.504850 &    19.201 &       0.089 &      r &       ZTF \\
           \hline
	\end{tabular}
\end{table}

\begin{table}
	\centering
	\caption{ZTF and ATLAS photometry.}
	\label{tab:photometry6}
	\begin{tabular}{lcccccc} 
          \hline
          MJD &
                Magnitude &  $\sigma_{\rm Magnitude}$& band & survey \\
          \hline
 59462.496401 &    19.191 &       0.112 &      r &       ZTF \\
 59462.498924 &    19.183 &       0.093 &      r &       ZTF \\
 59462.501435 &    19.243 &       0.128 &      r &       ZTF \\
 59462.503947 &    19.248 &       0.109 &      r &       ZTF \\
 59462.506470 &    19.257 &       0.122 &      r &       ZTF \\
 59467.580000 &    19.076 &       0.226 &      c &     ATLAS \\
 59469.475440 &    19.258 &       0.113 &      r &       ZTF \\
 59469.620000 &    19.160 &       0.330 &      o &     ATLAS \\
 59471.471319 &    19.304 &       0.159 &      r &       ZTF \\
 59472.451921 &    19.263 &       0.141 &      g &       ZTF \\
 59472.495880 &    19.296 &       0.118 &      r &       ZTF \\
 59474.449583 &    19.152 &       0.151 &      g &       ZTF \\
 59474.450058 &    19.076 &       0.113 &      g &       ZTF \\
 59477.468634 &    19.226 &       0.175 &      g &       ZTF \\
 59477.469109 &    19.297 &       0.169 &      g &       ZTF \\
 59477.494433 &    19.318 &       0.131 &      r &       ZTF \\
 59479.446412 &    19.556 &       0.239 &      r &       ZTF \\
 59479.447824 &    19.524 &       0.257 &      r &       ZTF \\
 59479.590000 &    19.112 &       0.395 &      o &     ATLAS \\
 59483.487095 &    19.561 &       0.174 &      r &       ZTF \\
 59485.570000 &    19.295 &       0.468 &      o &     ATLAS \\
 59488.498125 &    19.373 &       0.125 &      g &       ZTF \\
 59489.476007 &    19.516 &       0.145 &      r &       ZTF \\
 59489.487616 &    19.432 &       0.109 &      g &       ZTF \\
 59491.620000 &    19.368 &       0.299 &      o &     ATLAS \\
 59493.580000 &    19.596 &       0.439 &      o &     ATLAS \\
 59496.455255 &    19.498 &       0.133 &      r &       ZTF \\
 59496.463044 &    19.480 &       0.126 &      r &       ZTF \\
 59496.487535 &    19.484 &       0.168 &      g &       ZTF \\
 59500.444896 &    19.434 &       0.163 &      g &       ZTF \\
 59500.445370 &    19.305 &       0.116 &      g &       ZTF \\
 59500.487569 &    19.549 &       0.126 &      r &       ZTF \\
 59501.570000 &    19.642 &       0.390 &      o &     ATLAS \\
 59503.374618 &    19.645 &       0.232 &      r &       ZTF \\
 59503.423819 &    19.566 &       0.187 &      r &       ZTF \\
 59503.479329 &    19.326 &       0.094 &      g &       ZTF \\
 59503.497211 &    19.417 &       0.134 &      g &       ZTF \\
 59505.590000 &    19.691 &       0.564 &      o &     ATLAS \\
 59507.590000 &    19.485 &       0.470 &      o &     ATLAS \\
 59509.422280 &    19.331 &       0.170 &      g &       ZTF \\
 59509.422743 &    19.695 &       0.217 &      g &       ZTF \\
 59509.510000 &    19.631 &       0.870 &      o &     ATLAS \\
 59511.530000 &    19.227 &       3.074 &      o &     ATLAS \\
 59515.610000 &    19.396 &       0.820 &      o &     ATLAS \\
 59517.465984 &    19.582 &       0.109 &      g &       ZTF \\
 59517.503484 &    19.788 &       0.147 &      r &       ZTF \\
 59517.580000 &    20.063 &       0.511 &      o &     ATLAS \\
 59518.373831 &    19.391 &       0.090 &      g &       ZTF \\
 59518.491493 &    19.698 &       0.111 &      r &       ZTF \\
 59519.570000 &    19.999 &       0.411 &      o &     ATLAS \\
 59521.451019 &    19.695 &       0.108 &      r &       ZTF \\
 59521.500000 &    19.759 &       0.488 &      o &     ATLAS \\
 59522.456447 &    19.590 &       0.115 &      g &       ZTF \\
 59523.449456 &    19.519 &       0.091 &      g &       ZTF \\
 59523.499363 &    19.738 &       0.132 &      r &       ZTF \\
 59523.570000 &    19.479 &       0.262 &      c &     ATLAS \\
 59524.446979 &    19.733 &       0.123 &      r &       ZTF \\
 59524.508090 &    19.575 &       0.115 &      g &       ZTF \\
 59525.455880 &    19.787 &       0.105 &      r &       ZTF \\
 59525.466505 &    19.628 &       0.110 &      g &       ZTF \\
 59525.500000 &    19.574 &       0.360 &      o &     ATLAS \\
 59526.448993 &    19.576 &       0.094 &      g &       ZTF \\
             \hline
	\end{tabular}
\end{table}

\begin{table}
	\centering
	\caption{ZTF and ATLAS photometry.}
	\label{tab:photometry7}
	\begin{tabular}{lcccccc} 
          \hline
          MJD &
                Magnitude &  $\sigma_{\rm Magnitude}$& band & survey \\
          \hline
 59526.488171 &    19.911 &       0.162 &      r &       ZTF \\
 59527.373715 &    19.633 &       0.266 &      r &       ZTF \\
 59527.520000 &    19.704 &       0.293 &      c &     ATLAS \\
 59529.382893 &    19.640 &       0.142 &      g &       ZTF \\
 59529.406782 &    19.909 &       0.148 &      r &       ZTF \\
 59529.530000 &    20.207 &       0.510 &      o &     ATLAS \\
 59530.406505 &    19.607 &       0.209 &      g &       ZTF \\
 59530.443368 &    19.832 &       0.242 &      r &       ZTF \\
 59531.380162 &    19.550 &       0.121 &      g &       ZTF \\
 59531.421238 &    19.936 &       0.132 &      r &       ZTF \\
 59532.416273 &    19.646 &       0.115 &      g &       ZTF \\
 59532.436007 &    19.702 &       0.163 &      r &       ZTF \\
          \hline
	\end{tabular}
\end{table}


\bsp	
\label{lastpage}
\end{document}